\documentclass[sn-mathphys-num]{sn-jnl}% Math and Physical Sciences Numbered Reference Style 
\usepackage{graphicx}%
\usepackage{multirow}%
\usepackage{amsmath,amssymb,amsfonts}%
\usepackage{amsthm}%
\usepackage{mathrsfs}%
\usepackage[title]{appendix}%
\usepackage{xcolor}%
\usepackage{textcomp}%
\usepackage{manyfoot}%
\usepackage{booktabs}%
\usepackage{algorithm}%
\usepackage{algorithmicx}%
\usepackage{algpseudocode}%
\usepackage{listings}%
\usepackage{bm}
\theoremstyle{thmstyleone}%
\theoremstyle{thmstyletwo}%

\theoremstyle{thmstylethree}%

\usepackage[normalem]{ulem}
\newcommand{\MB}[1]{{\color{black} #1}}

\begin{document}

\title[Article Title]{Improving turbulence control through
explainable deep learning}

\author*[1,2]{\fnm{Miguel} \sur{Beneitez}}\email{miguel.beneitez@manchester.ac.uk}

\author[3,2]{\fnm{Andres} \sur{Cremades}}\email{ancrebo@upv.es}

\author[4]{\fnm{Federica} \sur{Tonti}}\email{ftonti@umich.edu}

\author[5]{\fnm{Luca} \sur{Guastoni}}\email{luca.guastoni@tum.de}

\author*[4,2]{\fnm{Ricardo} \sur{Vinuesa}}\email{rvinuesa@umich.edu}

\affil[1]{\orgdiv{Department of Mechanical and Aerospace Engineering}, \orgname{The University of Manchester}, \orgaddress{\city{Manchester}, \postcode{M13 9PT}, \country{UK}}}

\affil[2]{\orgdiv{FLOW, Engineering Mechanics}, \orgname{KTH Royal Institute of Technology}, \orgaddress{\city{Stockholm}, \postcode{SE-100 44}, \country{Sweden}}}

\affil[3]{\orgdiv{Instituto Universitario de Matemática Pura y Aplicada}, \orgname{Universitat Politécnica de Valéncia}, \orgaddress{\city{Valencia}, \country{Spain}}}

\affil[4]{\orgdiv{Department of Aerospace Engineering}, \orgname{University of Michigan}, \orgaddress{\street{Ann Arbor}, \city{MI 48109}, \country{USA}}}

\affil[5]{\orgdiv{School of Computation, Information and Technology}, \orgname{Technical University Munich}, \orgaddress{\street{85748 Garching}, \city{Munich}, \country{Germany}}}

%%==================================%%
%% Sample for unstructured abstract %%
%%==================================%%

\abstract{Turbulent-flow control aims to develop strategies that effectively manipulate fluid systems, such as the reduction of drag in transportation and enhancing energy efficiency, both critical steps towards reducing global CO$_2$ emissions. Deep reinforcement learning (DRL) offers novel tools to discover flow-control strategies, which we combine with our knowledge of the physics of turbulence. We integrate explainable deep learning (XDL) to objectively identify the coherent structures containing the most informative regions in the flow, with a DRL model trained to reduce them. The model trained with XDL targets the most relevant regions in the flow to sustain turbulence and produces a drag reduction which is higher than that of a model specifically trained to reduce the drag, resulting in a $18.1\%$ better net-energy saving. The XDL-based control remains the most effective control strategy when generalizing across Reynolds numbers and geometries. This demonstrates that combining DRL with XDL can produce causal control strategies that precisely target the most influential features of turbulence. By directly addressing the core mechanisms that sustain turbulence, our approach offers a powerful pathway towards its efficient control, which is a long-standing challenge in physics with profound implications for energy systems, climate modeling and aerodynamics.}

%\abstract{Turbulent-flow control aims to develop strategies that effectively manipulate fluid systems, such as the reduction of drag in transportation and enhancing energy efficiency, both critical steps towards reducing global CO$_2$ emissions. Deep reinforcement learning (DRL) offers novel tools to discover flow-control strategies, which we combine with our knowledge of the physics of turbulence. We integrate explainable deep learning (XDL) to objectively identify the coherent structures containing the most informative regions in the flow, with a DRL model trained to reduce them. The trained model targets the most relevant regions in the flow to sustain turbulence and produces a drag reduction which is higher than that of a model specifically trained to reduce the drag, while using only half its power consumption. Moreover, the XDL model results in a better drag reduction than other models focusing on specific classically identified coherent structures. This demonstrates that combining DRL with XDL can produce causal control strategies that precisely target the most influential features of turbulence. By directly addressing the core mechanisms that sustain turbulence, our approach offers a powerful pathway towards its efficient control, which is a long-standing challenge in physics with profound implications for energy systems, climate modeling and aerodynamics.}

\maketitle

Turbulence remains one of the last unsolved problems of classical physics~\cite{frisch1995turbulence}, defying a complete theoretical description despite its ubiquitous presence in natural~\cite{ramalingam1960convective} and engineering~\cite{spillane1967clear} flows. In fact, turbulent fluid flows are present in a myriad of industrial applications: from the food industry to the cooling of microprocessors, or harnessing wind power. Turbulence is also the primary cause of viscous drag with $30\%$ of the energy consumption worldwide being spent on overcoming drag in transportation~\cite{IEA2020}. Thus, improving the aerodynamic efficiency of terrestrial and airborne vehicles plays a pivotal role in the global reduction of CO$_2$ emissions and would have enormous economical consequences, since a reduction in drag achieving a fuel reduction on truck fleets of $5\%$ would correspond to an increase in profit of over $100\%$\cite{brunton2015closed}. Despite centuries of theoretical~\cite{marusic_science}, experimental~\cite{avila_science}, and computational~\cite{cardesa_science} progress since Leonardo da Vinci's first observations \cite{marusic2021leonardo}, many fundamental questions on the nature of turbulence remain unsolved~\cite{gibson1962spectra,argoul1989wavelet,greenstein1970superfluid}, preventing us from efficiently manipulating turbulent flows.

\begin{figure}
    \centering
    \includegraphics[width=0.8\linewidth]{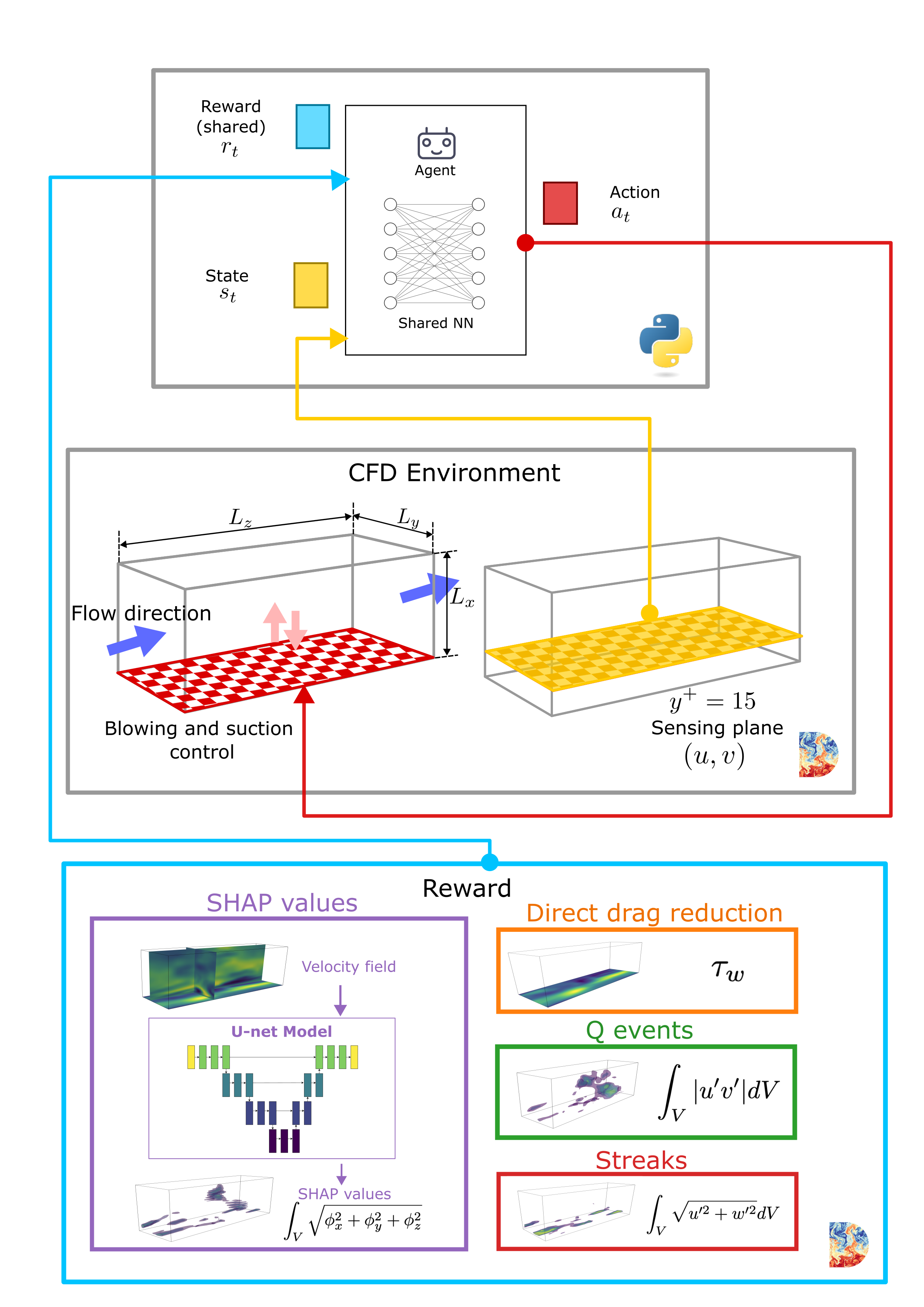}
    \caption{\textbf{The overall architecture of the deep reinforcement learning framework}. The scheme shows the communication between the involved parts. The top shows the information inputs to the DRL agent, \textit{i.e.} the reward ($r_t$) and state ($s_t$) at each time $t$, and outputs to the flow, \textit{i.e.} the action ($a_t$). In this multi-agent reinforcement-learning (MARL) framework, the agents cooperate featuring a shared neural network. The middle panel depicts the computational fluid dynamics (CFD) environment, from which each agent receives the information $(u,v)$ of the point above within a plane located at $y^+=15$. The actions are applied at the wall in the form of blowing and suction. At the bottom, we illustrate the different rewards considered. We note that the SHAP values are computed instantaneously from the velocity field through a pre-trained U-net model.}
    \label{fig:sketch}
\end{figure}

In this work, we incorporate the most recent progress in understanding the underlying dynamics of turbulence to design novel control strategies which reduce the friction produced by the flow. In the chaotic motion that characterizes wall-bounded turbulence, it is possible to identify persistent spatio-temporal patterns (coherent structures) that interact in the \textit{near-wall cycle of turbulence}~\cite{jimenez1999autonomous,jimenez2013near}. This cycle involves streamwise vortices generating elongated velocity perturbations (streaks), which subsequently experience instabilities and breakdown, generating new vortices and giving rise to a self-sustaining process (SSP)~\cite{waleffe1997self}. Existing approaches to drag reduction, such as opposition control~\cite{choi1994active}, seek to disrupt this SSP based on heuristic observations, for instance by targeting so-called ejection and sweep events~\cite{jimenez2018coherent}. Recently, the advent of new data-intensive techniques 
has expanded the available toolbox, with deep reinforcement learning (DRL) emerging as a particularly powerful approach to reveal new, more effective control strategies. DRL has demonstrated success across diverse physical domains, as broad as plasma fusion~\cite{seo2024avoiding}, optics~\cite{nousiainen2021adaptive}, and  fluid dynamics. Within this domain, DRL has been successfully applied to a wide range of control problems, including coordinated motion in fish schooling~\cite{gazzola2014reinforcement} and the control of separated flows~\cite{gautier2015closed,font2025deep}.

The framework for DRL control is illustrated in Fig. \ref{fig:sketch}: the DRL agent (or controller) receives partial observations $s_t$ of the system and is trained to generate actions $a_t$, which modify the turbulent flow in the effort to maximize the reward $r_t$. While previous DRL applications to turbulence control have targeted directly the reduction of turbulent drag~\cite{guastoni2023deep, Sonoda_Liu_Itoh_Hasegawa_2023}, our approach targets the SSP of near-wall turbulence, with rewards based on the agent's ability to manipulate turbulent coherent structures. 

We identify these structures using both conventional criteria (c.f. \cite{jimenez2018coherent} and references therein) and data-driven methods based on explainable deep learning (XDL)~\cite{cremades2024identifying}. Conventional coherent structures in turbulence, \textit{i.e.} Q events, streaks, and vortices, are characterized using instantenous velocity fluctuations and their derivatives, while data-driven structures are identified using Shapley additive explanation (SHAP) values~\cite{lundberg2017unified}, which are obtained through a game-theoretic method that calculates the importance of each input feature in a deep neural network. This idea has been successfully applied to explain relaminarization events in wall-bounded shear flows \cite{lellep2022interpreted}, where a XGBoost model trained to predict relaminarization events performed best when trained on features identified by the SHAP values. When applied to a U-net based on convolutional layers used for prediction of the future states of the flow, the SHAP values have successfully identified regions of importance in wall-bounded turbulence~\cite{cremades2024identifying}. 

The present study demonstrates how XDL can effectively be coupled with DRL to improve the performance of the learned policies compared with conventional physics-based rewards. We will firstly show the strength of the XDL and DRL combination on a two-dimensional illustrative problem, moving afterwards towards numerical simulations of turbulent flows. In the flow problem we will evaluate four reward strategies: (i) wall-shear stress reduction, (ii) Q-event reduction, (iii) streak reduction, and (iv) reduction of the SHAP values. The first reward is the most direct approach, rewards (ii) and (iii) target coherent structures traditionally studied in the turbulence literature, while (iv) is entirely data-driven yet physics-aware via XDL. To design the rewards, we choose appropriate instantaneous proxies to characterize the intensity of the turbulent coherent structures. In the SHAP-based reward, a U-net is used to predict the SHAP values from the instantaneous velocity field, as discussed in more detail in the Methods section. 

Our work presents an entirely data-driven method that is free from assumptions about which coherent structures are most relevant for turbulence, and yet is capable of identifying the physical processes governing wall-bounded turbulence dynamics. Should we consider an entirely different flow problem, the SHAP values would need to be recomputed. However, in the present work we show that predicting them through a U-net trained to predict SHAP values from an uncontrolled velocity field yields good results for control purposes. The implications of the present study extend beyond turbulence control, suggesting a powerful tool where XDL and DRL can uncover fundamental physical mechanisms that might otherwise remain out of reach using traditional approaches.

\section*{Targeting coherent structures for drag reduction}\label{sec2}

\begin{figure}
    \centering
    \includegraphics[width=1.0\linewidth]{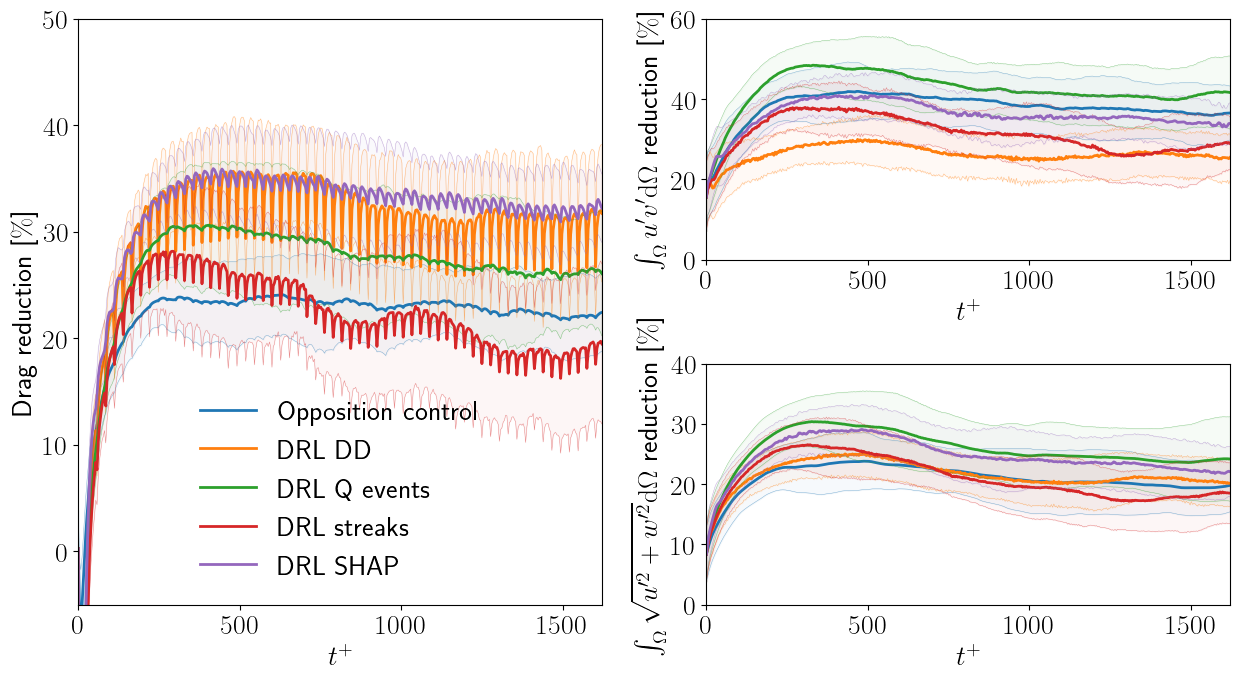}
    \caption{\textbf{Reduction of the quantities of interest with respect to the uncontrolled case}. (Left) Drag reduction, (top-right) proxy for Q-event reduction and (bottom-right) proxy for streak reduction. Results are averaged over the 150 initial conditions used for policy evaluation. Solid lines denote mean values and shaded regions indicate one standard deviation. Colors indicate different control strategies: opposition control (blue), DRL for direct drag reduction (orange), DRL for Q-event reduction (green), DRL for streak reduction (red), and DRL for SHAP reduction (purple). }
    \label{fig:channel_comp} 
\end{figure}

Our DRL set-up strategically identifies coherent structures to effectively influence the SSP of near-wall turbulence. This approach requires quantitative proxies to characterize the coherent structures in the flow. 
We define the instantaneous velocity vector $\mathbf{u}(x, y, z, t) = (u, v, w)$ along the streamwise, wall-normal and spanwise directions, where $t$ denotes time. Deviations from the instantaneous spatial mean averaged in the periodic directions are represented by primed variables: $u'(t,x,y,z) = u(t,x,y,z)-\iint u(t,x,y,z)\mathrm{d}x\mathrm{d}z/(L_xL_z)$ and $L_x$ and $L_z$ denote the domain sizes. We consider a turbulent open channel which can be characterized by the friction Reynolds number $Re_{\tau}=u_{\tau}h/\nu$, where $\nu$ denotes the fluid kinematic viscosity, $h$ the channel height, and $u_{\tau}=\sqrt{\tau_{w}/\rho}$ is the friction velocity (defined in terms of the wall-shear stress $\tau_w$ and the fluid density $\rho$). We consider $Re_{\tau}=180$ in our uncontrolled simulations. Quantities  non-dimensionalized with the viscous scales $u_{\tau}$ and $\nu$ are denoted by the superscript $``+"$. The DRL-discovered control strategies will be compared with the heuristic opposition control. Opposition control is a closed-loop control, where the control action is based on real-time flow-state sensing~\cite{choi1994active}, which reduces friction drag by using blowing and suction at the wall with a vertical velocity distribution defined as:
\begin{equation}
     v_\mathrm{wall}(x,z,t) = -v'(x,y_s,z,t).
\end{equation}
This control opposes the vertical velocity on a plane at a wall-normal distance $y_s$, with the aim of suppressing the streamwise vortices. The following control inputs are sampled at $y^+=15$: $v'(t,x,y^+=15,z)$ for opposition control and $\{u'(t,x,y^+=15,z),v'(t,x,y^+=15,z)\}$ for the DRL-based techniques. To quantify the total intensity of Q events we consider the Reynolds stress averaged over the domain volume: 
\begin{equation}
    \int_\Omega |(u(t,x,y,z)-U_T(y)) v(t,x,y,z)| \rm{d}\Omega,
\end{equation}
where $U_T$ is the long-time spatio-temporal average for the turbulent uncontrolled case and $\Omega$ denotes the volume. Following~\cite{kline1967structure}, we characterize the streak intensity as:
\begin{equation}
    \int_\Omega \sqrt{u'(t,x,y,z)^2+w'(t,x,y,z)^2} \rm{d}\Omega.
\end{equation}
Similarly, we define the presence of highly informative regions based on the SHAP values as:
\begin{equation}
    \int_\Omega \sqrt{\phi_x'(t,x,y,z)^2+\phi_y'(t,x,y,z)^2+\phi_z'(t,x,y,z)^2} \rm{d}\Omega,
\end{equation}
where $\bm{\phi}=(\phi_x,~\phi_y,~\phi_z)$ represent the SHAP vector field as detailed in the Methods section. Here, the SHAP values are inferred instantaneously from the velocity field by a U-net trained on the uncontrolled channel data (see the Methods section for more details). Note that these proxies only quantify the magnitude of the relevant coherent structure. We train our DRL agents to reduce these coherent structures without considering a priori wether they might be beneficial or not for drag reduction.

We consider two channel domain sizes in our study, one for training and one for testing. The DRL agents are trained in a small channel configuration (SCC), the so-called minimal flow unit~\cite{jimenez1991minimal}, which is computationally affordable and supports a single SSP. The learnt policy is then tested in a large channel configuration (LCC) where many SSPs are present and interact. The number of agents is always kept equal to the number of gridpoints at the wall in our simulations, i.e. the evaluation of the policy in the LCC is achieved by replicating the learnt control. Here we discuss the performance in the larger channel, while the details of the computational settings and the SCC results are reported in the Methods section and in the Supplementary material, respectively.

The discovered policies exhibit significant differences between the various proposed rewards. Fig. \ref{fig:channel_comp} shows various performance metrics averaged over 150 different initial conditions. Our results exhibit a significant drag reduction: approximately \MB{$31.9\%$ when using DRL aiming specifically at direct drag (DD) reduction, $26.9\%$ when targeting Q events, and $20.9\%$ when targeting streaks. However, all these approaches show a worse performance than that of the SHAP-based policy, which exhibits $33.7\%$ drag reduction: 1.8 percentage points more than DD-based control, 6.8 percentage points more than Q-event-based control, and 12.8 percentage points more than streak-based control}. This finding is particularly remarkable given that all policies were initially trained in a small channel where performance differences were small (see the Methods section and Supplementary material). We note the strong periodicity of the time signal, which could be smoothed by considering a shorter actuation \MB{$dt_c$} (not shown). This is further illustrated in the zero-shot generalization for $Re_{\tau}=550$, where the actuation frequency remains the same in inner units, but is increased in outer units. It might appear counter-intuitive that a model trained to reduce the SHAP values outperforms a model trained for direct drag reduction when reducing the drag. However, the system under consideration is not entirely Markovian due to partial observability and the SHAP values can be seen as sophisticated reward shaping. A more detailed discussion using an illustrative navigation problem can be found in the Supplementary Material.

\MB{The SHAP values do not inherently explain causal relations. In this application they are obtained as explanations of a deep-learning model where the input is a fully observed system and the output also corresponds to the same fully observed system, evolved continuously under a differentiable deep-learning model. We can then expect the SHAP values to implicitly capture the underlying causal relations embedded in the model, acting as causal proxies, as supported by recent evidence \cite{cremades2024classically,molina2026unveiling} where the SHAP values are able to correctly identify the causally relevant structures in the Q2/Q4 pairs proposed in \cite{jimenez2018coherent}. We consider the SHAP framework to be an empirically promising approach to represent certain causal relations in turbulence, although further efforts are required to fully understand how these causal relationships are captured by the SHAP values.}

We obtain further insights into the dynamic effect of the DRL-discovered control strategies when comparing the averaged Reynolds stresses and streak intensity in the various controlled cases, as shown in Fig.~\ref{fig:channel_comp} (right). Note that the DRL approach targeting Q events exhibits the best performance reducing both Q events and streaks. Interestingly, the SHAP-based policy yields the highest drag reduction, despite not being particularly effective at reducing Q events or streaks, highlighting the fact that these classically studied structures are not the most important to target when reducing drag. The effectiveness of the SHAP-based policy stems from its ability to identify the most informative flow regions to predict future turbulence fluctuations in an objective manner. The DRL approach targeting SHAP values manipulates the most relevant structures for the flow evolution and the SSP of near-wall turbulence. By targeting these regions, the DRL control learns to hamper turbulence-supporting mechanisms in a causally informed manner, unlike the more simplistic approach of targeting the classically-studied coherent structures. These findings confirm that the traditional coherent structures provide an incomplete picture of the mechanisms supporting turbulence, and show that XDL complements and expands our understanding of wall-bounded turbulence~\cite{cremades2024classically}.

\begin{figure}
    \centering
    \includegraphics[width=1.0\linewidth]{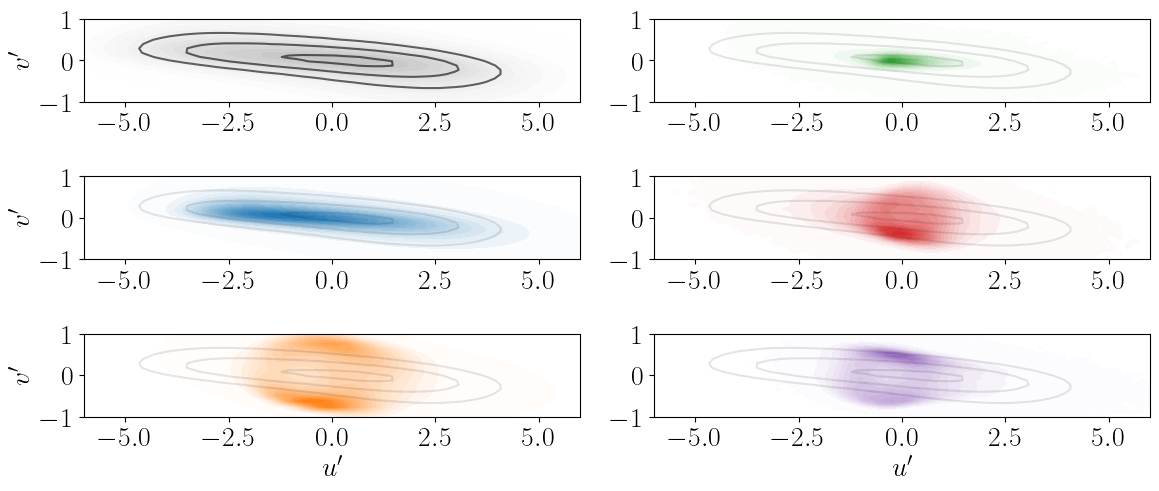}
    \caption{\textbf{Joint probability density function of the streamwise and wall-normal fluctuations at $y^+=15$ for the different control strategies}. Results are averaged over the 150 initial conditions used for policy evaluation for: uncontrolled flow (black), opposition control (blue), DRL for DD (orange), DRL for Q events (green), DRL for streaks (red), and DRL for SHAP values (purple). Grey lines in all the panels show the uncontrolled case for comparison.}
    \label{fig:quadrant}
\end{figure}

Fig.~\ref{fig:quadrant} shows the joint probability density function of streamwise ($u'$) and wall-normal ($v'$) perturbations on the sensing plane $y^+=15$ evaluated on 150 different trajectories. Note that the predominance of ejections and sweeps in the uncontrolled case is consistent with the wall-bounded turbulence literature~\cite{lozano2012three} and the results from opposition control and DRL for drag reduction agree with our previous results~\cite{guastoni2023deep}. Furthermore, the approaches based on classical structures produce distinctly different fluctuations intensities: Q-event-based control substantially reduces the perturbation intensities in both stream- and wall-normal directions, while the streak-based control decreases streamwise perturbations but increases wall-normal perturbations, emphasizing the \MB{negative values of the wall-normal fluctuations}. In contrast, our SHAP-based control leads to a quite even distribution of the four quadrants, with a slight preference towards positive streamwise \MB{wall-normal} fluctuations. Note that we are exclusively reporting the effect of the control on the velocity perturbations of the flow. The effect of each particular control on the drag reduction is highlighted in Fig.~\ref{fig:channel_comp}. Additional insights into the effects of the various control strategies are included in the Supplementary material, which includes flow visualizations and spectral analyses.

\begin{figure}
    \centering
    \includegraphics[width=1.0\linewidth]{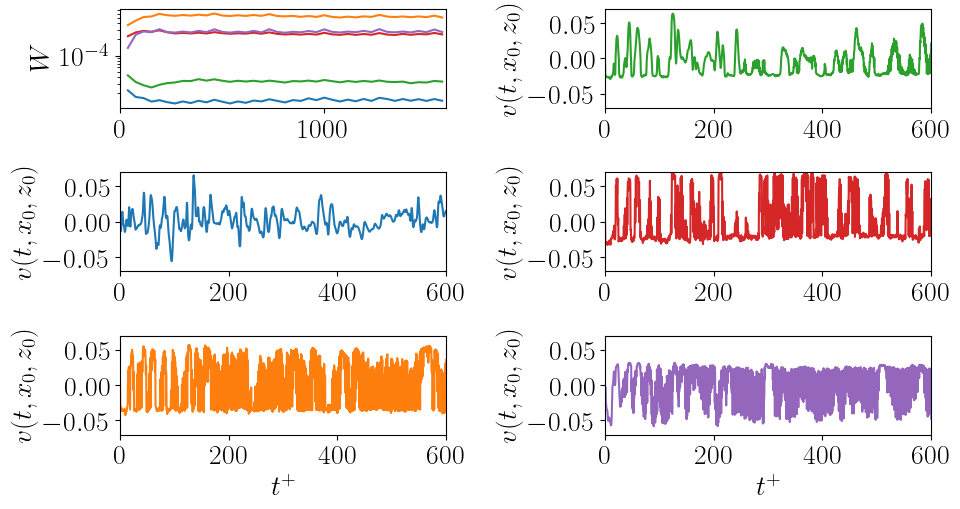}
    \caption{\textbf{Power input and time signal of a single controller for the various control strategies.} (Top left) Total power input averaged over the 150 initial conditions used for policy evaluation. (Rest of panels) Sample time signals for a single actuator in: opposition control (blue), DRL for drag reduction (orange), DRL for Q events reduction (green), DRL for streak reduction (red), and DRL for SHAP reduction (purple). The actuator location is the same in all the cases: $(x_0,y_0,z_0)=(0,0,0)$.}
    \label{fig:agents}
\end{figure}

An additional advantage of our XDL approach is observed when analyzing control dynamics and energy requirements. The total power input of the control is defined as:
\begin{equation}
    W = \frac{1}{2}S_1v||v_{\text{wall}}||^2 + S_2pv \label{eq:power_input}
\end{equation}
for each agent, where the first term denotes the kinematic power input, and  $S_{1,2}=0$ if $v\leq0$ (or $pv\leq0$), and $S_{1,2}=1$ if $pv>0$ following \cite{hasegawa2011dissimilar,Sonoda_Liu_Itoh_Hasegawa_2023} to ensure that no energy recovery from the pressure is considered. Fig.~\ref{fig:agents} shows sample actuation signals from the different control strategies, along with their associated power input requirements. Direct drag reduction control frequently oscillates between the minimum and maximum allowed control values in a ``bang-bang" pattern, resulting in high power input. The SHAP-based control and streak-based control operate more efficiently, rarely reaching maximum or minimum control values and exhibiting fewer extreme actuation patterns. This translates to considerable input power savings: the SHAP-based approach requires only half the power input compared with the DD control while delivering better drag reduction. The Q-event-based control, while even more energy-efficient, yields significantly lower drag reduction than the SHAP-based approach, as discussed above.

The net energy saving is defined as:
\begin{equation}
    S = \frac{c_{f,\text{uncontrolled}}- (c_{f}+W ) }{c_{f,\text{uncontrolled}}}, \label{eq:net_energy_saving}
\end{equation}
where the skin-friction coefficient is $c_f = 2\tau_w/(\rho U_b^2)$ and $U_b$ denotes the bulk velocity. This quantity, which discounts the power required by the control from the drag reduction, further emphasizes the advantage of SHAP-based control over the other methods. In particular, \MB{DRL for SHAP reduction yields $30.6\%$ net-energy saving compared with $25.9\%$ achieved by the DRL for direct drag reduction method (4.7 percentage points better), $26.4\%$ achieved by the DRL for Q events method (4.2 percentage points better), $22.7\%$ achieved by opposition control (7.9 percentage points better), and $18\%$ achieved by the DRL for streaks (12.6 percentage points better)}. While the power input is generally small compared to the skin-friction coefficient ($W\sim \mathcal{O}(10^{-4})$ and $c_f\sim \mathcal{O}(10^{-3})$), these efficiency gains become significant in large-scale applications requiring thousands of controllers. Furthermore, equation \eqref{eq:power_input} underestimates the power input in an actual experimental setting~\cite{kametani2011direct}, where the advantage of the SHAP-based approach compared with the DRL for direct drag reduction would become even more pronounced.

\begin{figure}
    \centering
    \includegraphics[width=0.7\linewidth]{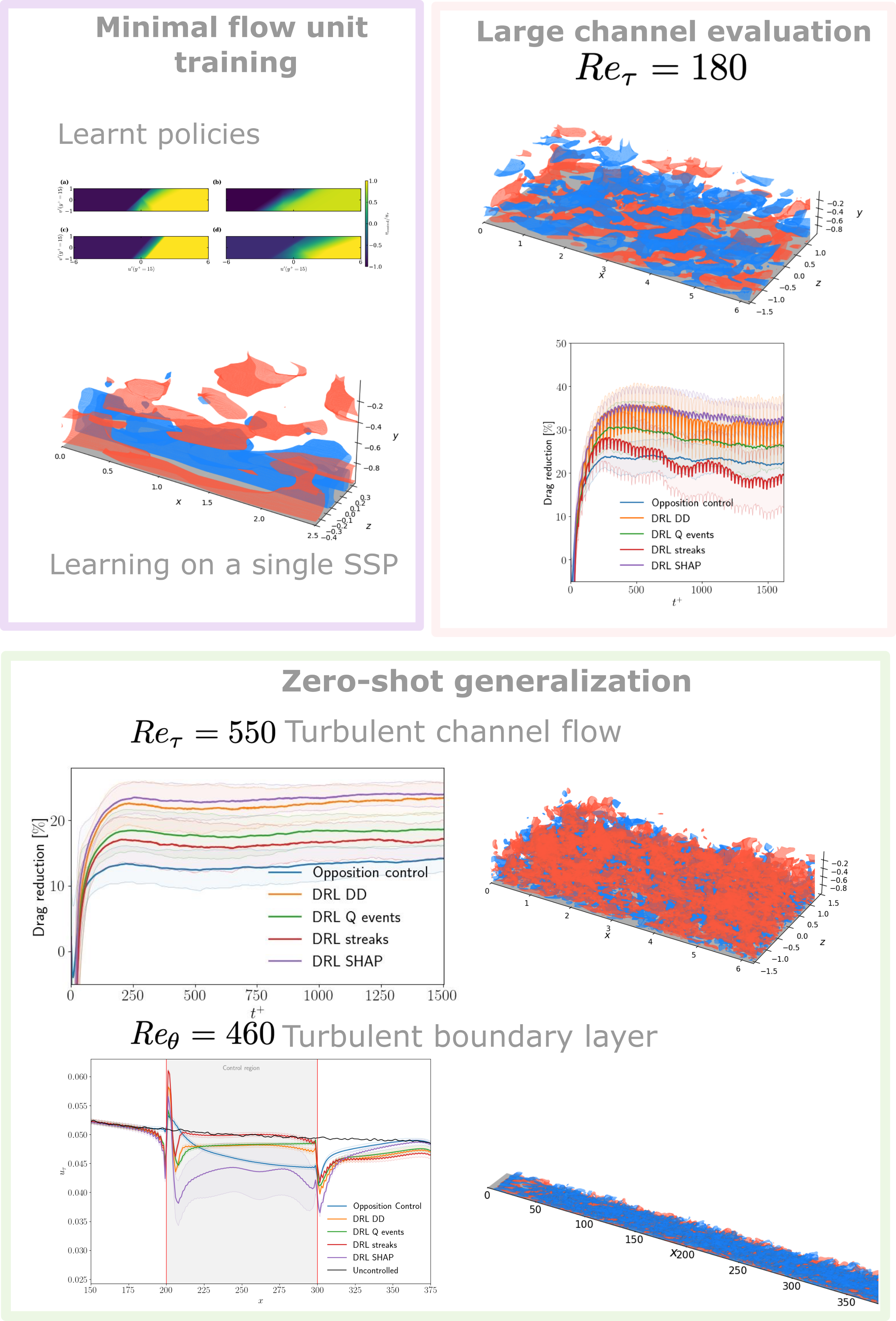}
    \caption{\textbf{Illustration of the Z
    zero-shot generalization process.} (Top left) Illustration of the learnt policies and training in SCC at $\textit{Re}_{\tau}=180$, red and blue contours indicate illustrative values of $v$. (Top right) Evaluation of the policy on a large channel (LCC) at $\textit{Re}_{\tau}=180$, red and blue contours indicate illustrative values of $v$. (Bottom) Results from the zero-shot application of the policies learnt on the SCC at $\textit{Re}_{\tau}=180$ on a turbulent channel flow at $\textit{Re}_{\tau}=550$ and a turbulent boundary layer with a $\textit{Re}_{\theta}=460$, red and blue contours indicate illustrative values of $w$. These flow cases are detailed in the Methods section and 3D visualizations for the uncontrolled cases are added to illustrate the flow complexity. In all cases the SHAP-based policy is the best performing one.}
    \label{fig:zero_shot} 
\end{figure}

\section*{Zero-shot generalization}
Zero-shot deployment of a DRL has proved to be successful even in complex geometries, such as the flow around wings~\cite{wang2026physics}. We further assess the drag-reduction strategies discovered in this work through a zero-shot generalization to two other flow cases: (i) a channel flow at $\textit{Re}_{\tau}=550$ and (ii) zero-pressure-gradient turbulent boundary layer (ZPGTBL).

Firstly, we consider the turbulent flow in the LCC at a $\textit{Re}_{\tau}=550$ instead of the training one of $\textit{Re}_{\tau}=180$. Our DRL policy is deployed in this configuration by keeping the same set-up in inner units, i.e. sensing plane remains at the same $y^+=15$, that the amplitude of the control remains $u_{\tau}$, and the control frequency remains at the same \MB{$dt_c^+\approx 0.54 $}, for consistency, we keep the observations normalized with the original $u_{\tau}$ from training. The results for the drag reduction averaged over 150 different initial conditions corresponding to the various DRL policies shown in Fig. \ref{fig:zero_shot}. We observe that while all policies degrade with respect to the $\textit{Re}_{\tau}=180$ case, the SHAP-based policy still outperforms all other policies resulting in a drag reduction of $23.5\%$, which is $4.3\%$ better (1 percentage point) than the DD-based control, $29.8\%$ better (5.4 percentage points) than the Q events-based control, $42.4\%$ better (7 percentage points) than the streaks-based control and $77.4\%$ better (10.25 percentage points) than opposition control. This shows that the SHAP-based control effectively generalises well beyond its training configuration (as long as the SHAP structures observed in the flow remain similar).

Secondly, we consider a ZPGTBL with $\textit{Re}_{\delta_0^*}=450$ at the inflow. We apply the control between $x\in[200,300]$ downstream of the inflow, where the turbulence fully developed \cite{schlatter2009turbulent,guastoni2025fully} (more details can be found in the Methods section). In this case, since the flow is spatially developing, we use the uncontrolled streamwise-averaged quantities in inner units within the control region for our setup. Fig.~\ref{fig:zero_shot} shows the spatial distribution of the $u_{\tau}$ along the boundary layer averaged in time, in the spanwise direction, and over 150 different initial conditions. The results show that in this configuration the DRL trained to minimize the SHAP values in the flow yields the lowest $u_{\tau}$. The drag reduction obtained within the control region is $32\%$ for the SHAP-based control, this is 23.9 percentage points better than the DD-based control, 24.3 percentage points better than the Q events-based control and 16.6 percentage points better than opposition control. We note that the streaks-based control only produces $0.05\%$ drag reduction in this case, being mostly inefficient. It is worth noting that opposition control is better than all of the DRL-discovered control strategies, except for the SHAP-based control. 

\MB{Our empirical evidence shows that a zero-shot transfer of the SHAP-based policy is successful as long as the SHAP values are similar between cases. We observed a worsening performance when applying the existing control policy to flow configurations with significantly different SHAP distributions. Nevertheless, we also observed that a short transfer learning applied to the previously converged DRL policy on the new case resulted in SHAP-based control strategies which outperformed all other control methods in all set-ups explored. We have effectively checked that the drag reduction from the SHAP-based policy reported here is robust to changes in the sensing plane and the time interval used to calculate the SHAP values (see Supplementary Information).}

These zero-shot generalizations are applied to flow configurations significantly different from the training set-up to illustrate the flexibility and robustness achieved through XDL for turbulence control. 

\section*{Conclusions}

This study demonstrates a novel approach to turbulence control by integrating explainable deep learning (XDL) with deep reinforcement learning (DRL) using the Shapley additive explanations (SHAP) values. The SHAP values identify the most important coherent structures in the flow and they are capitalized on to discover control strategies for wall-bounded turbulence. By training DRL models in a minimal flow configuration with a single self-sustaining process, we successfully developed control strategies which can be deployed accross Reynolds numbers and geometries through zero-shot generalization on a channel at $\textit{Re}_{\tau}=550$ and a zero-pressure-gradient turbulent boundary layer at $\textit{Re}_{\theta}=460$.

Our analysis reveals several critical insights. While the DRL approach targeting Q events excels at reducing the classical Q events and streaks, it falls short in reducing those structures truly relevant for drag reduction. In contrast, the SHAP-based approach, relying on the SHAP values, yields a more effective control strategy, achieving $33.7\%$ drag reduction. This represents a $5.6\%$ improvement over the DRL approach directly targeting drag reduction, a $25.3\%$ improvement over Q-event-based control, and a $61.2\%$ improvement over the streak-based control. Furthermore, the SHAP-based control only requires half of the power input compared with the DRL approach targeting drag reduction. This yields even more significant results, where the SHAP-based control achieves a net energy saving of $30.6\%$, which is $15.9\%$ better than controlling Q events, $18.1\%$ better than controlling the drag, $34.8\%$ better than opposition control, and $70\%$ better than controlling the streaks. Additionally, we show that the SHAP-based control is still the best performing policy in a turbulent channel flow at $\textit{Re}_{\tau}=550$ and for the ZPGTBL at $\textit{Re}_{\theta}=460$ without any retraining. The boundary layer case is particularly relevant, since the SHAP-based policy is the only control technique outperforming opposition control without any specific retraining.

It is also interesting to note that in the SHAP-based approach all input information comes from a single plane parallel to the wall and that all training is performed with just a single SSP, and consequently, the amount of experiences observed in training is limited. This approach demonstrates the potential of XDL in identifying critical flow mechanisms, where the proposed method is capable of capturing the causal relationships between flow structures, rather than targeting individual coherent structures in isolation. 

The framework introduced in this work is a promising entry point for the powerful and unique combination of DRL and XDL, where there is still a wide range of possible improvements such as: usage of a convolutional neural network in the DRL policy (instead of a multilayer perceptron), training using volumetric data instead of a single plane, or transfer learning from the small training domain to the large channel one. The use of XDL for DRL can have a great impact in a wide range of applications as it can be applied directly on experimental data \cite{cremades2024identifying}, avoiding the ever growing requirements for computational power and high-resolution simulation. The SHAP reward introduced in this work provides new insights into the most critical mechanisms of turbulence and provides new paths to tame and harness the power of turbulent flows in many industrial (e.g. aerodynamics and food production), biological (e.g. active matter) and medical applications (e.g. personalised medicine). 

\section*{Methods}\label{sec11}

\subsection*{Computational set-up}

As introduced in the main text, we consider a turbulent open-channel flow of height $h$, driven by a time-varying pressure gradient that maintains a constant mass flux. The spatial coordinates are $x$, $y$, and $z$ in the streamwise, wall-normal, and spanwise directions, respectively. We solve the Navier--Stokes equations, which are made non-dimensional with the open-channel height $h$ and the laminar centerline velocity $U_{cl}$, and read:
\begin{align}
    \partial_t \mathbf{u} + (\mathbf{u} \cdot \nabla)\mathbf{u} &= -\nabla p + \frac{1}{\textit{Re}}\Delta \mathbf{u},\\
    \nabla \cdot \mathbf{u} &= 0. 
\end{align}
The flow field is entirely described by the velocity vector $\mathbf{u}(x, y, z, t) = (u, v, w)$. Direct numerical simulations (DNS) for training are carried out in a computational domain of dimensions $[L_x,L_y,L_z]=[2.67,1,0.8]$, i.e. the small channel configuration (SCC), and the evaluation of the resulting policies is performed in a box of size $[L_x,L_y,L_z]=[2\pi,1,\pi]$, i.e the large channel configuration (LCC). We consider Dirichlet boundary conditions at the lower wall and symmetry boundary conditions at the upper boundary:
\begin{align}
    u(y=0)=w(y=0)=0,\ v(y=0)=v_{bc},\\ \partial_y u(y=1)=\partial_y w(y=1)=0,\ v(y=1)=0.
\end{align}
The uncontrolled turbulent flow is characterized by the friction Reynolds number, which is set to $Re_{\tau}=180$.

To perform our DNS we use the generic partial differential equation (PDE) solver Dedalus~\cite{burns2020dedalus}. This solver uses a pseudo-spectral method, where solutions are expanded into Fourier series in the homogeneous ($x$ and $z$) directions and into Chebyshev polynomials into the inhomogeneous (wall-normal, $y$) direction. We expand solutions of the SCC into $[N_x,N_y,N_z]=[16,64,16]$ modes and we timestep the simulations using a third-order four-stage diagonally implicit Runge--Kutta and explicit Runge--Kutta (DIRK+ERK) scheme~\cite{ascher1997stabilization} with a constant time step \MB{$dt^+ \approx 0.039$ ($dt=0.005$)}. These computational parameters are common in the literature for the Reynolds number considered in this study \cite{guastoni2023deep}. The LCC simulations are performed expanding the solutions into $[N_x,N_y,N_z]=[64,64,32]$ modes using the same time-stepping scheme and time step as in the SCC case. These configurations (SCC and LCC) are used to perform long simulations ($t^+>20,000$) which form the dataset for training (and testing) required for the U-net models and the DRL framework. The set-up for the zero-shot generalization for a $\textit{Re}_{\tau}$=550 uses the same geometry as the LCC, and is expanded into $[N_x,N_y,N_z]=[512,256,64]$ modes using the same timestepping scheme as the SCC with a constant time step \MB{$dt=0.001$.}

\subsubsection*{Turbulent boundary layer set-up}

We consider a zero-pressure-gradient turbulent boundary layer (ZPGTBL) developing on a flat plate. In this configuration we use the same coordinates as in the SCC but $x$ is measured from the domain inflow, located at a distance $x_0$ from the leading edge. A local Reynolds number can be defined as $\textit{Re}_{\delta^*}=U_{\infty}\delta^*(x)/\nu$, where $U_{\infty}$ denotes the freestream velocity  and $\delta^*(x)=\int_0^{L_y}(1-u/U_{\infty})dy$ is the displacement thickness. The equations are nondimensionalized by $U_{\infty}$ and $\delta^*_0$. We consider a computational domain of dimensions \MB{$[L_x,L_z,L_y]=[500,30,60]$} where $\textit{Re}_{\delta_0^*}=450$ is set at the inflow. We use the generic solver Dedalus~\cite{burns2020dedalus}, which allows only for one inhomogeneous direction, and project our solution onto Fourier modes in the stream- and spanwise directions and Chebyshev polynomials in the wall-normal direction. We use $[N_x,N_z,N_y]=[512,128,256]$ modes and to satisfy the periodicity in the $x$-direction despite the spatial development we use the fringe method detailed in \cite{chevalier2007simson}. We consider a fringe of length $x_{\text{fringe}}=135$ at the end of the domain, with a rise region of $100$, a fall region of $30$ and a maximum amplitude of $0.8$. Turbulence is triggered by a random trip volume forcing applied at $x_{\text{trip}}=10$ from the start of the computational domain using the forcing detailed in \cite{chevalier2007simson}. This results in a measured $Re_{\theta}=460$ at the start of the control region, where $\theta=\int_0^{L_y}u/U_{\infty}(1-u/U_{\infty})dy$

\subsection*{Reinforcement learning configuration and training}

Setting up a control environment requires the consideration of several elements, such as the position of the sensing plane, the control frequency and amplitude, as well as any parameters affecting the computation of the rewards. We set up our training configuration following the best practices in the literature. The distance from the wall of the sensing plane has been considered in \cite{vinuesainfluence}, concluding that $y^+=15$ results in the best performance, both for DRL for direct drag reduction and for opposition control. The frequency and intensity of the control are chosen in agreement with previous works \cite{guastoni2023deep,Sonoda_Liu_Itoh_Hasegawa_2023}. The time delay used for the calculation of the SHAP values has been considered in \MB{Refs.~}\cite{cremades2024identifying,cremades2024classically} concluding that $\Delta t^+=\{1,2,5,10\}$ do not significantly affect the identified coherent structures based on the SHAP values. More details about the calculation of the SHAP values can be found in the section below.

As briefly described in the introduction, a DRL set-up consists of two main elements that interact with each other: the environment, i.e. the system where we input actions, obtain rewards, and observe states, and the agent, i.e. the controller in charge of deciding the actions to take based on inputs from the environment. In the present work, as shown in Fig.~\ref{fig:sketch}, the environment corresponds to a numerical simulation performed using the Dedalus codebase~\cite{burns2020dedalus}, which accounts for the main computational cost in the DRL set-up. To address this, the Dedalus code is parallelized over several processors using the message-passing interface (MPI). 

As mentioned, the DRL setup requires the choice of an agent, which can be classified as model-based or model-free. Model-based agents are based on a model of the dynamics of the environment, while model-free methods optimize the agent
in a trial-an-error process carrying out a number of episodes. In the present work we choose a model-free agent. Such agents can be further split into policy-gradient, value-function, and a combination of policy-gradient and value-function algorithms dubbed actor-critic algorithms. A policy-gradient algorithm is based on the parametrization of the agent and its optimization to maximize cumulative rewards. Value-function algorithms, in contrast, attempt to estimate the cumulative reward given a state (state value function $V^{\pi}(s)$) or a
state-action couple (action value function $Q^{\pi}(s, a)$), where:
\begin{align}
    V^{\pi}(s) &= \mathbb{E}_{\pi}\left[ r|s\right], \\
    Q^{\pi}(s,a) &= \mathbb{E}_{\pi}\left[ r|s,a\right], 
\end{align}
with $\mathbb{E}$ denoting the expectation. Moreover, model-free agents can be classified as on-policy or off-policy. In on-policy methods, the optimization and policy update are based on the learning generated by the current agent. Off-policy algorithms use a replay buffer to store trajectories generated with previous policies, and these are used to perform the policy update. In the present work we have chosen the model-free, off-policy TD3 algorithm~\cite{fujimoto2018addressing} as implemented in the Stable Baselines 3 library \cite{stable-baselines3}. The combination of the Dedalus codebase with Stable Baselines 3 is particularly convenient since both of them are Python based. Note that this is a robust state-of-the-art DRL algorithm suitable for continuous control problems~\cite{fujimoto2018addressing}. The reinforcement learning side of the set-up is much less computationally demanding and can be efficiently trained on a single central-processing unit (CPU). 

In the physical configuration for the SCC described above, we defined the same number of controllers (agents) as grid points (modes) in the periodic directions, such that each agent sees only the state $(u,v)$ located at $y^+=15$ right above. The spatial average of the state is removed at each step, i.e. whenever information is communicated to the agent to avoid biasing the observations when evolving the flow between the controlled and uncontrolled cases (note that this is only required when information is passed to the agent). Consequently, each actuator is defined as a multi-agent reinforcement-learning pseudo-environment, resulting in $N_x \times N_z = 256$ independent trajectories that the agent uses during the optimization step. The actions taken by the agents correspond to blowing and suction, i.e. positive/negative values of the vertical velocity at the wall, and they are limited to the range $[-u_{\tau}, u_{\tau}]$. In order to ensure that no net mass is introduced by the controllers, we enforce a zero-mean action over all agents. This is done by removing the spatial average over all actuators before the control is applied in the DNS.  The control interacts with the DNS every \MB{$dt_c^+\approx 0.54$ ($dt_c=0.07)$} and each episode consists of $3,000$ interactions, in agreement with prior studies \cite{guastoni2023deep}. Between consecutive agent interactions, the instantaneous values of the reward are calculated and stored at each DNS step. These stored values are then averaged to produce a single reward value.

We train each model using 6 different initial conditions in the SCC. At the start of each episode a random number generator chooses which initial condition is used for that particular trajectory. This initial condition is then evolved for 3000 interactions between the agent and the flow field. The actuation frequency is set to \MB{$dt_c =0.07$} and the timestep of the simulation is set to \MB{$dt=0.005$}. The reward is collected at each timestep and averaged before sharing it with the agent every 14 timesteps in the DNS. Moreover, the policy is updated every 30 interactions between the agent and the DNS. Since the TD3 algorithm uses a deterministic policy in training, we introduce noise with amplitude $0.1u_{\tau}$ to favor exploration. During training in the SCC, relaminarization of the flow can be observed and in such cases we terminate the episodes once the flow is deemed to approach the laminar state sufficiently, i.e. once the drag reduction is over $55\%$. This is done to avoid providing the model with very large rewards for actions which no longer affect the flow as it is on its path to relaminarization. The model is evaluated every 5 episodes on an additional initial condition not seen in training for evaluation. The best performing model on the evaluated initial condition is then stored. Although  several evaluation episodes could be considered, this process becomes expensive and evaluating on a single initial condition proved to be sufficient to identify effective DRL models. Moreover, we perform several training runs to explore a wider range of control strategies. The Supplementary Material shows the reward metrics during sample training runs for the different rewards considered. It can be observed that typically the model improves quickly at the start of the process and then discovers many similarly effective policies during training. Throughout the training, the model passes through parameter regions where better policies are found, but also where it unlearns. Each 100 episodes corresponds to roughly 96 hours of computing using four AMD Ryzen 9 7950X CPUs.

The best policy for each of the four DRL rewards in the SCC is tested in 50 initial conditions randomly sampled from our database of $30,000$ instantaneous fields and unseen during training as reported in the Supplemetary material. The results show that all DRL policies outperform opposition control and for $t^+>500$ they achieve drag reductions of: $36.5\%$ for the DD reduction reward, $36.2\%$ for the Q-event-based reward, $34.4\%$ for the streak-based reward, and $37.4\%$ for the SHAP-based reward. It can also be observed that in this small channel, the SHAP-based reward produces the best performing policy in terms of reducing the Q events and streaks.

The training in the small channel results in a deterministic policy $a_t=\mu(s_t)$ where $a_t$ denotes the action and $s_t$ denotes the state. The deterministic policy in this case is a mapping $f:U\times V \rightarrow A$ where $U$ denotes all the possible values of $u'(x,y^+=15)$ (fluctuations in the streamwise direction), $V$ denotes all the possible values of $v'(x,y^+=15)$ (fluctuations  in the wall-normal direction) and $A$ denotes all the possible values of the action. Since our policy is a non-linear mapping it can be evaluated in any system independently of the number of agents, i.e. each agent has access to the same policy and selects an action accordingly. 

\subsection*{Calculation of the XDL-based reward}
Defining a criterion for calculating the reward requires determining the regions of the flow with a higher impact on its evolution. In order to calculate these high-importance regions, an explainable-deep-learning (XDL) approach is employed. This approach is based on three steps: i) predict a future state of the flow through a deep-learning model, ii) calculate the importance of each grid point in the prediction of flow in a future time and iii) train a model that predicts the field of importance (SHAP values) from the original velocity fluctuation field.

For the first stage, a U-net architecture~\cite{ronneberger2015}, $f_u$, is employed to predict the velocity fluctuation field in a future time step ($\mathbf{u}^p_{t+\Delta t}$) with $\Delta t^+=\Delta t\, u_\tau^2/\nu\approx 5$, from the true velocity field at the present time ($\mathbf{u}_{t}$): $\mathbf{u}^p_{t+\Delta t} = f_u\left(\textbf{u}_{t}\right)$. Both input and output fields have a size of $16\times 64 \times 16$ grid points in the streamwise, wall-normal and spanwise directions, respectively. In addition, a periodic padding of $3$ is added in the streamwise and spanwise directions. The U-net has approximately $1.5$ million parameters, distributed on $4$ levels in the U-net, with $16$, $32$, $64$, and $128$ filters and a kernel of size $3\times 3$ from the first to the last level. Each level is composed of two blocks (3D convolutional layer+batch normalization+activation function)  and one block in the decoder. The encoder levels are connected by average poolings with size $2$, while in the decoder, the levels are connected with transposed convolutions with $16$, $32$, and $64$ filters and kernel of size $3\times 3$ from top to bottom level. Each level of the encoder and the decoder are connected by concatenating the output of the encoder level with the output of the transposed convolution. The U-net is trained on a database of $30,000$ instantaneous flow fields, using $20\%$ for validation, until the prediction error is approximately $1\%$.

Once the model for predicting the velocity is trained, the importance of each grid point for the accuracy of the prediction is evaluated. For this reason, the model $f_u$ is modified in order to calculate the mean-squared error of the predictions:

\begin{equation}
\label{eq:mse_model}
{\rm MSE}_{t+\Delta t} = f_{\rm MSE}\left(\mathbf{u}_{t}\right)=\frac{1}{N_xN_yN_z}\sum_{i_x=1}^{N_x}\sum_{i_y=1}^{N_y}\sum_{i_z=1}^{N_z}\left(\mathbf{u}_{t+\Delta t}-f_u\left(\mathbf{u}_{t}\right)\right)^2\rm{.}
\end{equation}

Then, the importance of each grid point is calculated using additive-feature-attribution methods~\cite{lundberg2017unified}, which substitute the deep-learning model $f_{\rm MSE}$ by a surrogate linear model $g_{\rm MSE}$ defined as:

\begin{equation}
\label{eq:mse_shap}
   {\rm MSE}_{t+\Delta t} = f_{\rm MSE}\left(\mathbf{u}_{t}\right) \approx g_{\rm MSE}\left(z_{ji}\right) = \phi_0 + \sum_{j\in(u,v,w)}\sum_{i=0}^N\phi_{ji} z_{ji}\rm{,}
\end{equation}

\noindent where the importance, or Shapley additive explanations (SHAP) values, of each single grid point $i$ for the component $j$ of the velocity is defined by $\phi_{ji}$ and $N$ is the total number of grid points in a single snapshot. The parameter $z_{ji}$ is a boolean value which is $0$ or $1$ in case of removing or including the information of the velocity component, $j$, in this specific grid point, $i$. The expected mean-squared error of the prediction when all the grid points are removed is defined by $\phi_0$. Note that the linear model presented in equation (\ref{eq:mse_shap}) is a local approximation of the MSE, and thus, it is updated for every snapshot. However, the computational cost of these values increases exponentially with the number of parameters, i.e. as $2^N$~\citep{jia2019}. To reduce the computational cost of the calculations, the prior knowledge of the mathematical definition of the architecture and the neurons can be exploited~\cite{cremades2025}. In the present work, the gradient-SHAP algorithm, an additive-feature-attribution method for differentiable models based on the expected-gradients method~\cite{erion2021}, is used to simplify the SHAP-value calculation as follows:

\begin{equation}
    \label{eq:expected_gradient}
    \phi_{ji}\left(u_{t}\right) = \mathbb{E}_{u_{\rm{ref}},\alpha\sim U\left( 0,1\right)}\left[\left(u_{t_{ji}}-u_{t_{ji}}\right)\frac{\partial f_{\rm MSE}\left(u_{\rm{ref}}+\alpha\left(u_{t}-u_{\rm{ref}}\right)\right)}{\partial u_{t_{ji}}}\right]\text{,}
\end{equation}

\noindent where $u_{\rm{ref}}$ is the reference velocity field. For this analysis, the instantaneous spatially-averaged mean flow is used as a reference as the instantaneous mean value of the velocity is non-informative for the evaluation of the error of the model.

After the SHAP values of the $30,000$ instantaneous flow fields are calculated, a new U-net is trained to predict the SHAP values from the present velocity fluctuation field: $\bm{\phi}=f_\phi\left(\mathbf{u}_{t}\right)$. The architecture of this model is identical to that of the previous U-net ($f_u$), using in this case the instantaneous velocity field as input and the SHAP-value field as output. A database comprising the previous $30,000$ instantaneous flow fields is used to train the model, reserving $20\%$ for validation, until the prediction error is lower than $0.5\%$. Finally, the SHAP-value field $\bm{\phi}$ is predicted in order to compute the reward of the DRL in the simulation loop with a low computational effort compared to that of explicitly computing the SHAP values, making this step over 200 times faster. 

In the present setup, we use the aforementioned U-net without retraining to predict the SHAP values from full instantaneous velocity fields, even while control is applied. Although this might modify the SHAP values associated with the flow field under consideration, these values are computed with respect to an instantaneous spatial average. Therefore, suppressing SHAP values is expected to reduce fluctuations over the instantaneous mean flow, which on its own is unable to sustain turbulence. This further strengthens the robustness of the results: The SHAP values obtained for an uncontrolled turbulent flow yield good control strategies even when the underlying flow state is modified by a varying boundary condition at the wall.

While \citet{fryer2021shapley} and \citet{huang2024failings} raised significant concerns regarding the use of SHAP values for feature attribution, particularly their instability under correlation and tendency to assign importance to logically irrelevant features, these issues are mitigated in the present study due to the nature of turbulent flows. Dependencies in such physical systems are predominantly local, the risk of misleading global correlations is minimized. Furthermore, the inherent noise in feature attribution is suppressed by exploiting domain periodicity. By exploiting  these physical constraints, we ensure that SHAP-based policies remain effective and provide a faithful representation of the underlying dynamics despite the theoretical approximation limits of the Shapley framework, as previously demonstrated for multiple causal systems in the supplementary material of \citet{cremades2024classically}, where the SHAP values detected causal relationships in mediator, cofounder and collider systems. In fact, the SHAP values provide a sensitivity analysis for a deep-learning model. In general, this correlation does not imply causation. However, when the model is trained on time-resolved, high-quality  data such as the temporal evolution of a turbulent flow the resulting SHAP analysis reflects the causal dynamics of the flow.

%--------------------------------------------------------------------------------------------

\section*{\MB{Data availability}}

\MB{After acceptance, the data used in this study will be made available open access in the following repository: \url{https://github.com/VinuesaLAB-AI}}

\section*{\MB{Code availability}}

\MB{After acceptance, all the codes used in this study will be made available open access in the following repository: \url{https://github.com/VinuesaLAB-AI}}

\backmatter

\bmhead{Acknowledgements}

The deep-learning-model training was enabled by resources provided by the National Academic Infrastructure for Supercomputing in Sweden (NAISS) at Dardel (PDC) and Alvis (CS3E) and by the University of Michigan. M.B. and R.V. acknowledge partial financial support from Digital Futures. \MB{R.V. acknowledges financial support from the University of Michigan and} ERC grant no. ‘2021-CoG-101043998, DEEPCONTROL. Views and opinions expressed are those of the authors only and do not necessarily reflect those of the European Union or the European Research Council. Neither the European Union nor the granting authority can be held responsible for them.

\bmhead{Author contribution}
Beneitez, M.: Methodology, Software, Validation, Investigation, Writing - Original Draft, Visualization. Cremades, A.: Methodology, Software, Writing - Original Draft. \MB{F. Tonti: Methodology, Software, Validation, Writing - Revised Draft}. Guastoni, L.: Methodology, Software, Writing -
Original Draft. Vinuesa, R.: Conceptualization, Project definition, Methodology, Resources, Writing - Original Draft, Supervision, Project administration, Funding acquisition.

\begin{appendices}

\end{appendices}

%%===========================================================================================%%
%% If you are submitting to one of the Nature Portfolio journals, using the eJP submission   %%
%% system, please include the references within the manuscript file itself. You may do this  %%
%% by copying the reference list from your .bbl file, paste it into the main manuscript .tex %%
%% file, and delete the associated \verb+\bibliography+ commands.                            %%
%%===========================================================================================%%

\clearpage

%%==================================%%
%%      Supplementary Material      %%
%%==================================%%

\setcounter{figure}{0}
\renewcommand{\thefigure}{S\arabic{figure}}
\setcounter{equation}{0}
\renewcommand{\theequation}{S\arabic{equation}}
\setcounter{section}{0}
\renewcommand{\thesection}{S\arabic{section}}

\begin{center}
{\Large \textbf{Supplementary Material -- Improving turbulence control through explainable deep learning}}
\end{center}

\vspace{1em}

\section*{SCC results}
\begin{figure}
    \centering
    \includegraphics[width=1.0\linewidth]{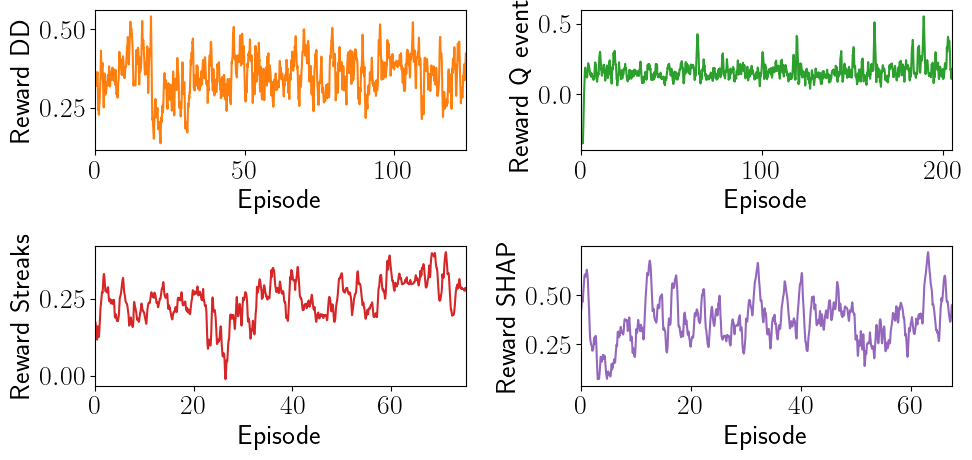}
    \caption{\textbf{Sample training runs for the DRL models in the SCC}. Reward for the DRL training is measured as reduction of the quantity of interest with respect to the uncontrolled case for various control strategies. Orange: DRL for reduction of the wall shear stress (i.e. direct drag reduction); Green: DRL for reduction of Q events; Red: DRL for reduction of streaks; Purple: DRL for reduction of SHAP values.}
    \label{fig:training}
\end{figure}

\begin{figure}
    \centering
    \includegraphics[width=1.0\linewidth]{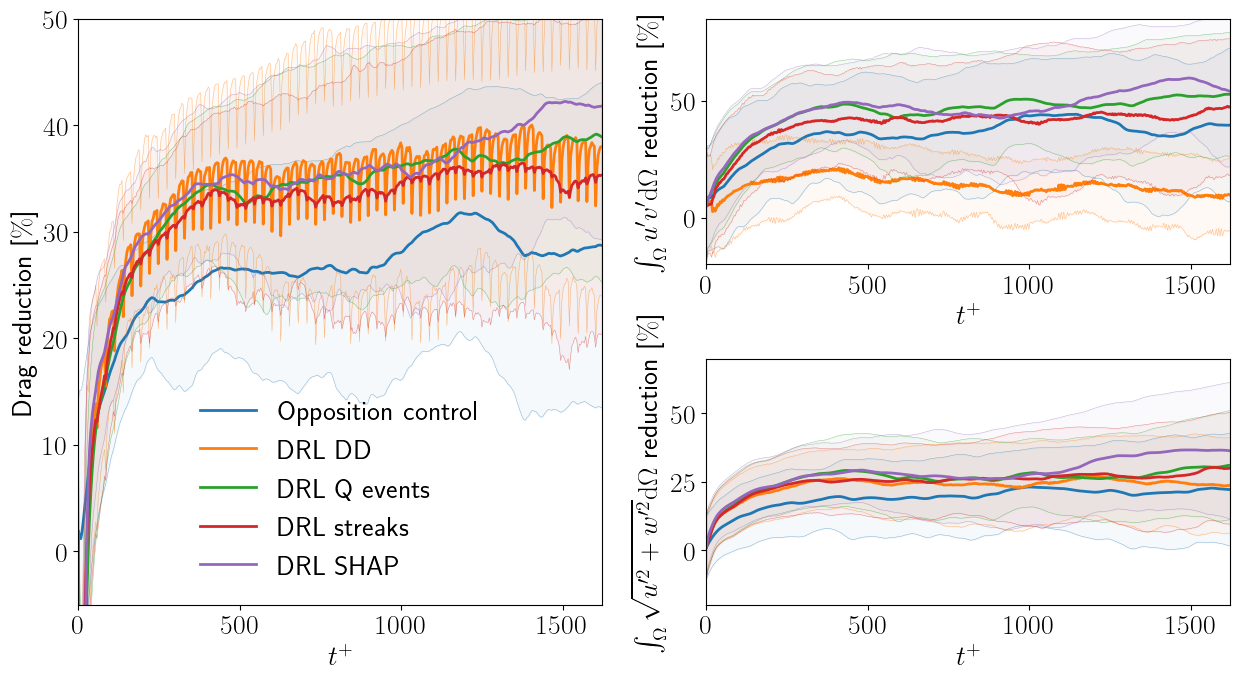}
    \caption{\textbf{Reduction of the quantities of interest with respect to the uncontrolled case in the SCC}. (Left) Drag reduction, (top-right) proxy for Q-event reduction and (bottom-right) proxy for streak reduction. Results are averaged over the 50 initial conditions used for policy evaluation. Solid lines denote mean values and shaded regions indicate one standard deviation. Colors indicate different control strategies: opposition control (blue), DRL for direct drag reduction (orange), DRL for Q-event reduction (green), DRL for streak reduction (red), and DRL for SHAP reduction (purple).}
    \label{fig:channel_comp_small} 
\end{figure}

In this section we provide additional results regarding the SCC described in the main paper, including training metrics and the evaluation of the DRL-discovered policies. As mentioned in the main paper, we train all models using 6 different initial conditions. At the start of each episode a random number generator chooses which initial condition is used for that particular trajectory. We employ the TD3 algorithm with amplitude $0.1u_{\tau}$ to favor exploration. As mentioned in the main paper, several training runs are carried out for each of the rewards considered (direct drag reduction, Q events, streaks and SHAP values) and the best performing model over all runs and episodes is chosen for each reward. The reward communicated to the agent is normalized as follows:
\begin{equation}
r_t = 1-\frac{\gamma}{\gamma_{\text{uncontrolled}}},
\end{equation}
where $\gamma$ accounts for any of the proxies described in the main paper. This expression aims to keep the maximum reward bounded from above by 1, which is recommended for DRL applications. Fig.~\ref{fig:training} shows the reward at each training episode for various sample training runs. This figure shows that typically the model improves quickly at the start of the process and then discovers many similarly effective policies during training. This behaviour is in agreement with the literature \cite{guastoni2023deep,Sonoda_Liu_Itoh_Hasegawa_2023} and is consistent over several training runs (not shown). Fig.~\ref{fig:channel_comp_small} shows various performance metrics averaged over 50 different initial conditions unseen in training for the SCC. The results show that in this configuration the SHAP-based policy outperforms all others in every metric, achieving a drag reduction of $37.4\%$, a Q-event reduction of $49.8\%$, and a streak reduction of $30.6\%$. In comparison, opposition control achieves a drag reduction of $28.1\%$, the DD-based policy achieves a drag reduction of $36.5\%$, the Q-event-based policy achieves a drag reduction of $36.2\%$, and the streak-based policy achieves a drag reduction of $34.4\%$.

\begin{figure}
    \centering
    \includegraphics[width=1.0\linewidth]{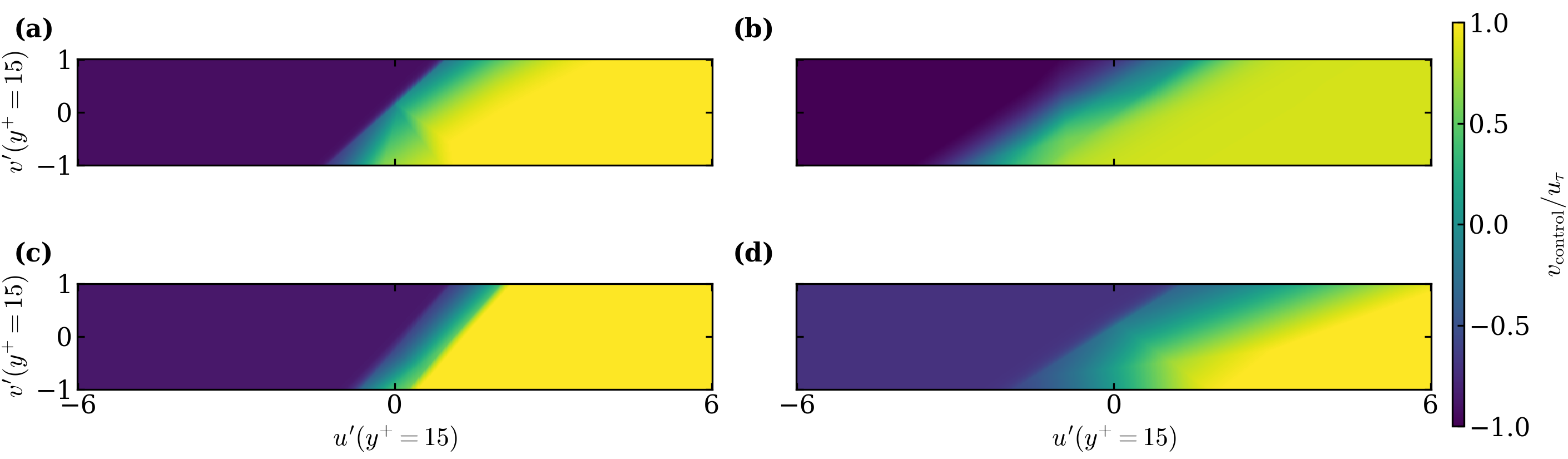}
    \caption{\textbf{Learnt policies for the considered rewards in the SCC}. (a) Direct drag reduction, (b) SHAP values, (c) proxy for streak reduction and (d) proxy for Q events. $x$ and $y$ axis show the input variables to the agent and colormap indicates the normalised output for the control action. All learnt policies behave similarly for larger values of $u'(y^+=15)$, but differ significantly for small values of that input. }
    \label{fig:learnt_policies} 
\end{figure}

Fig.~\ref{fig:learnt_policies} examines the policies resulting from the DRL training.  The policies provide a map from the agent inputs, i.e. $u'(y^+=15)$ and $v'(y^+=15)$, to the control action $v_{\text{control}}$ for the various choices of rewards. We observe that the main difference between policies are related to the response to values of $u'(y^+=15)$, for larger values of this input all policies behave in the same way, providing high blowing when the streamwise velocity perturbations are large and high suction when this input is significantly negative. However, the behaviour for input values of $u'(y^+=15)$ between $[-1,1]$ is significantly different. For the direct drag reduction policy (Fig.~\ref{fig:learnt_policies} (a)) we observe a sharp change between blowing and suction control, very similar to the results reported by \cite{guastoni2023deep, Sonoda_Liu_Itoh_Hasegawa_2023}. The SHAP-based policy instead shows a transition between blowing and suction characterized by a smoother gradient, rather than a sharp change. The streak-based policy exhibits  the most abrupt blowing-to-suction transition, which also happens at higher values of $u'(y^+=15)$. A smoother transition between blowing and suction occurs for Q events-based policy, which also shows the most nonlinear mapping between inputs and outputs.

\section*{Further analysis of the LCC}

%---------------- Quantitative data
\begin{figure}
    \centering
    \includegraphics[width=0.49\linewidth]{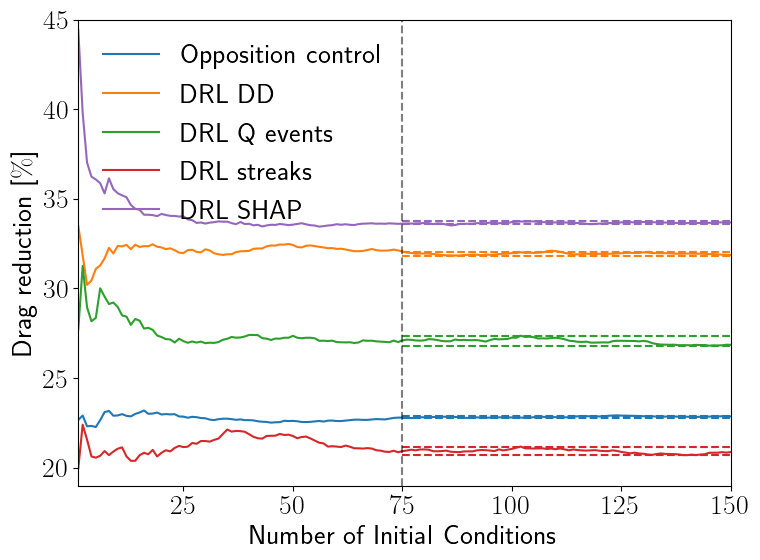}
    \includegraphics[width=0.49\linewidth]{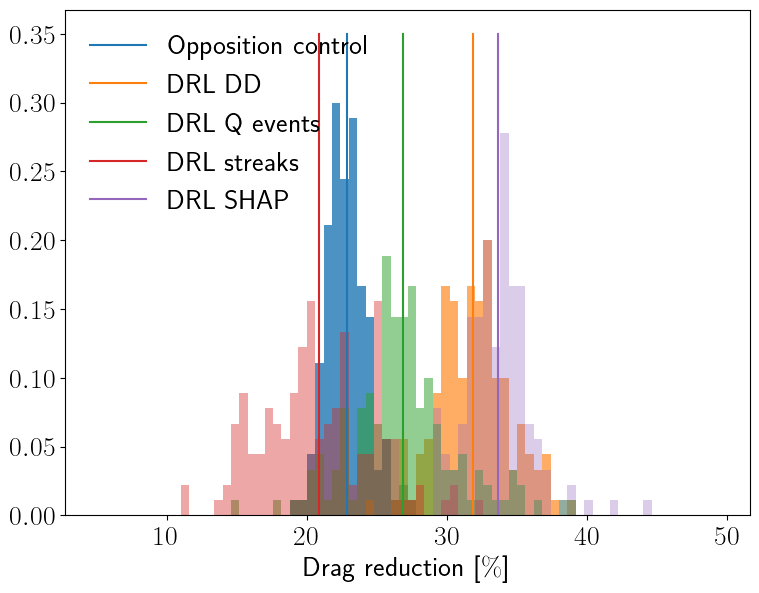}
    \caption{\textbf{Statistical assessment of the control policies in the LCC}. (Left) Average drag reduction for each policy based on the number of initial conditions evaluated. Dashed lines indicate a 95$\%$ confidence interval. The value of the drag reduction can be considered converged beyond 75 initial conditions. (Right) Histogram showing the drag reduction for each initial condition considered for each control strategy. Vertical lines indicate the mean values.}
    \label{fig:quant_policy}
\end{figure}

%---------------- Visualizations
\begin{figure}
    \centering
    \includegraphics[width=1.0\linewidth]{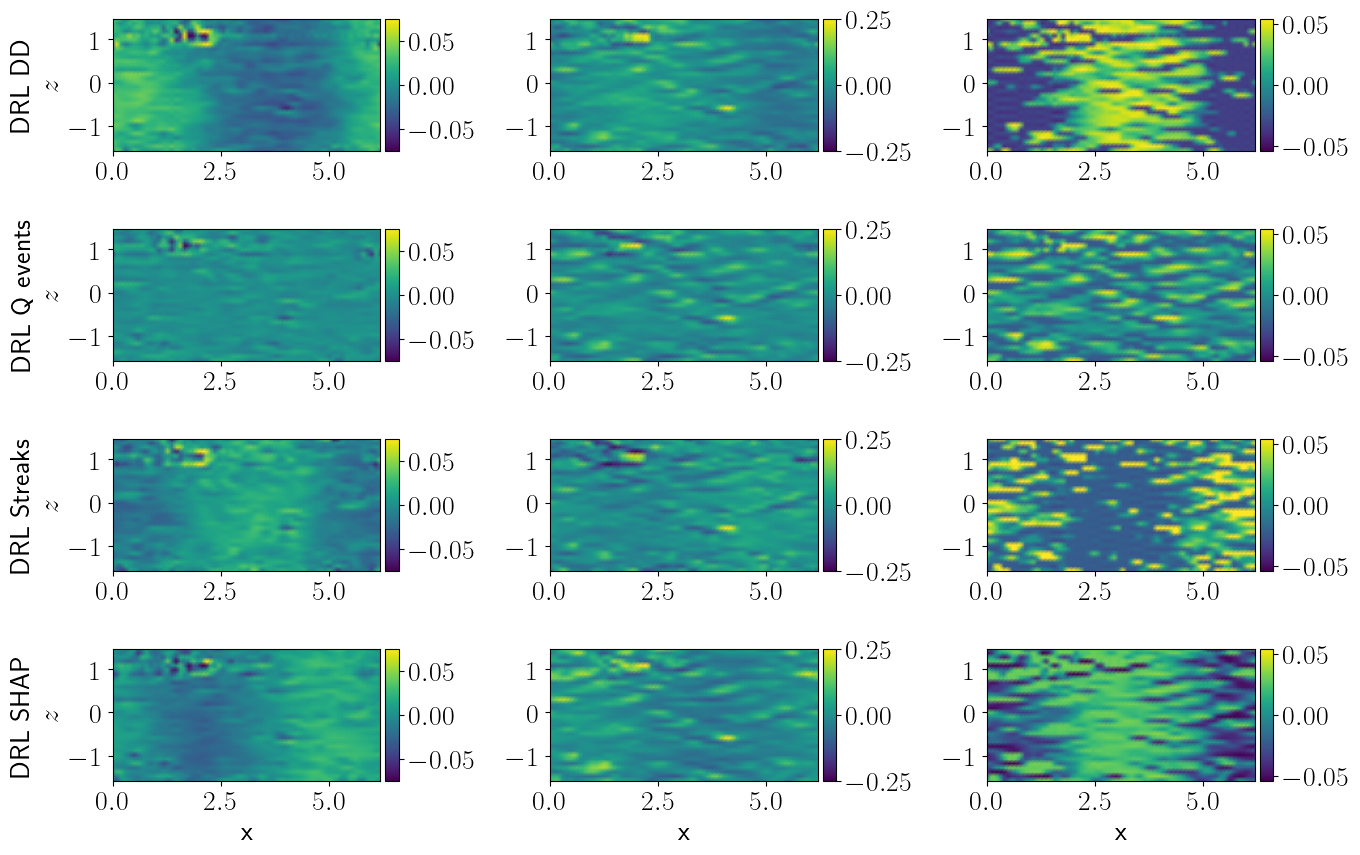}
    \caption{\textbf{Sample visualizations of the DRL inputs and the discovered control strategies at $t^+=100$ for the LCC}. (Left) Wall-normal velocity fluctuation on the sensing plane $v'(t^+=100,x,y^+=15,z)$, (middle) streamwise velocity component on the sensing plane $u'(t^+=100,x,y^+=15,z)$, right) wall-normal velocity at the wall induced by the control $v(t^+=100,x,y^+=0,z)$.}
    \label{fig:planes_100}
\end{figure}

\begin{figure}
    \centering
    \includegraphics[width=1.0\linewidth]{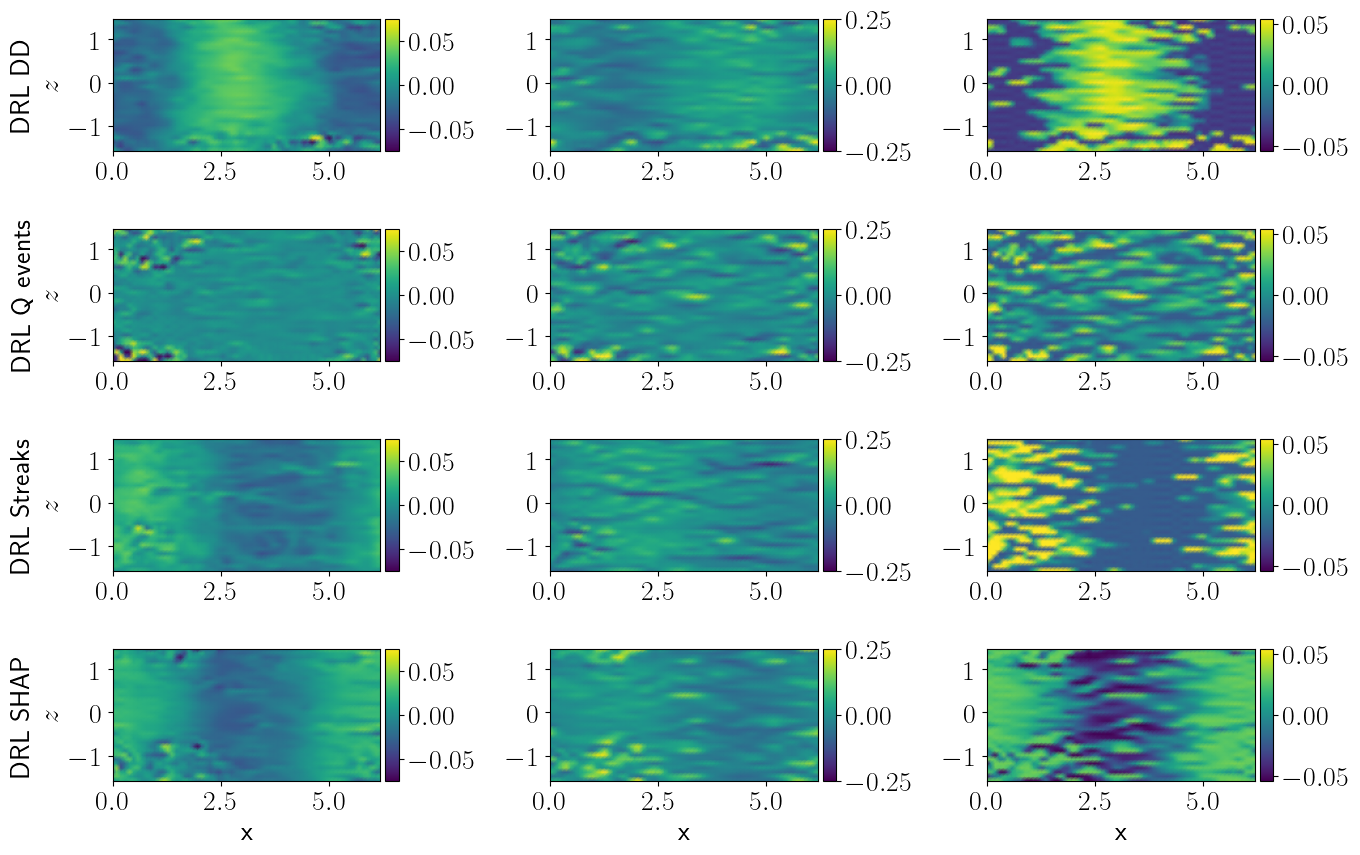}
    \caption{\textbf{Sample visualizations of the DRL inputs and the discovered control strategies at $t^+=1500$ for the LCC}. (Left) Wall-normal velocity fluctuation on the sensing plane $v'(t^+=1500,x,y^+=15,z)$, (middle) streamwise velocity component on the sensing plane $u'(t^+=1500,x,y^+=15,z)$, (right) wall-normal velocity at the wall induced by the control $v(t^+=1500,x,y^+=0,z)$.}
    \label{fig:planes_1500}
\end{figure}
\begin{figure}
\centering
\begin{tabular}{ccc}
 Uncontrolled &
  \raisebox{-.5\height}{\includegraphics[width=0.42\linewidth]{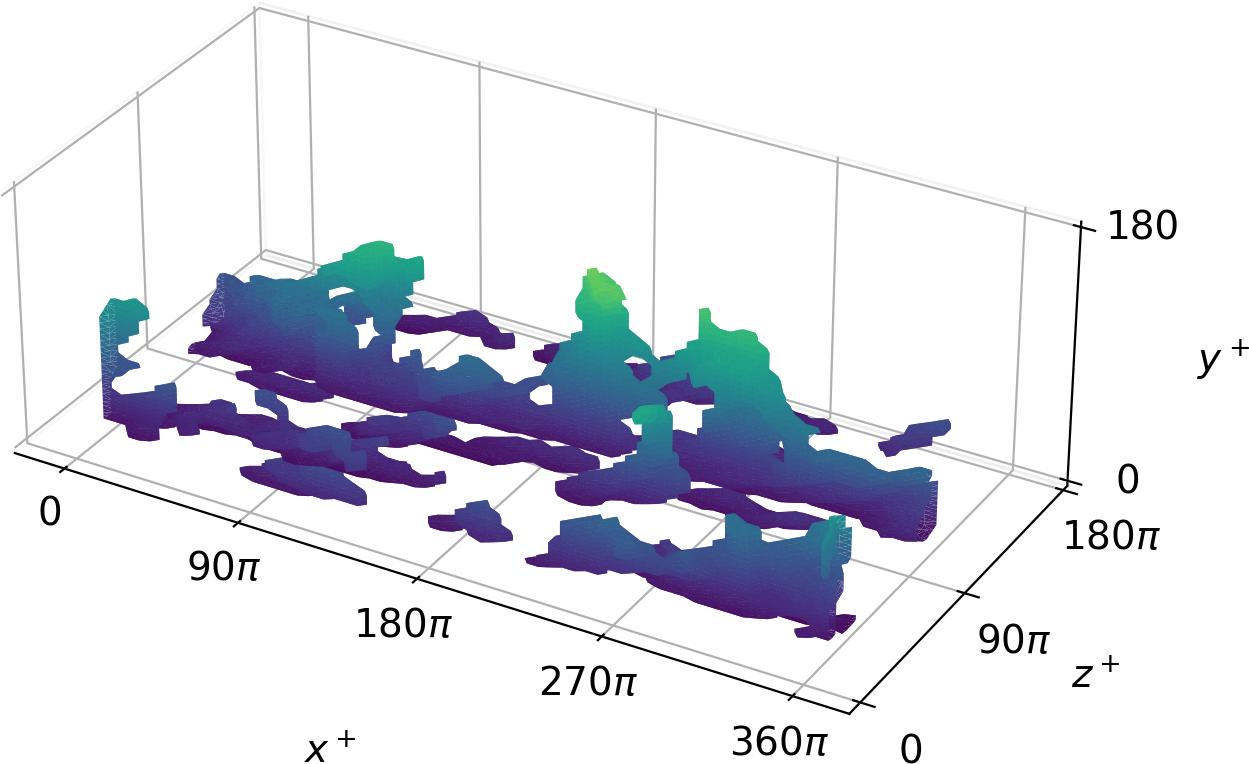}} & 
  \raisebox{-.5\height}{\includegraphics[width=0.42\linewidth]{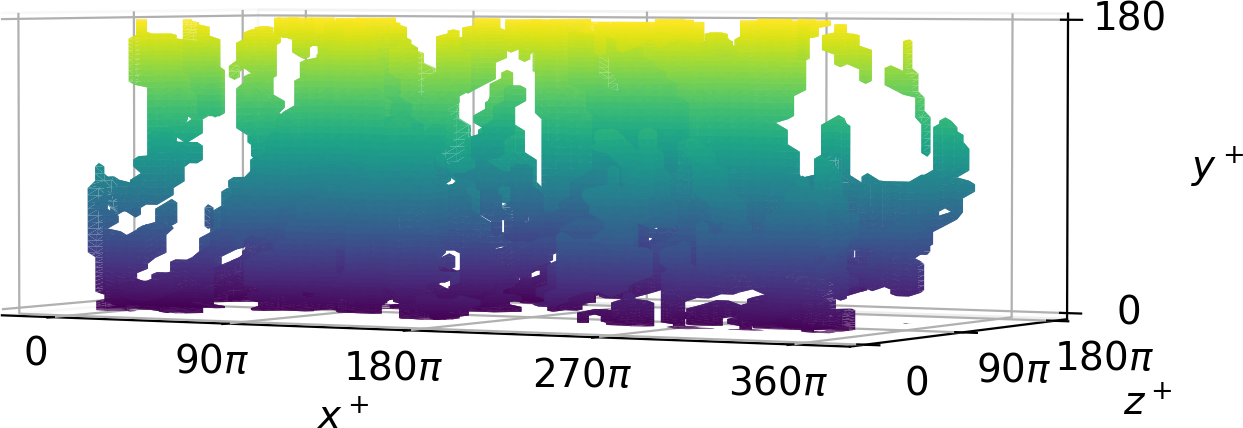}} \\

  DRL DD &
  \raisebox{-.5\height}{\includegraphics[width=0.42\linewidth]{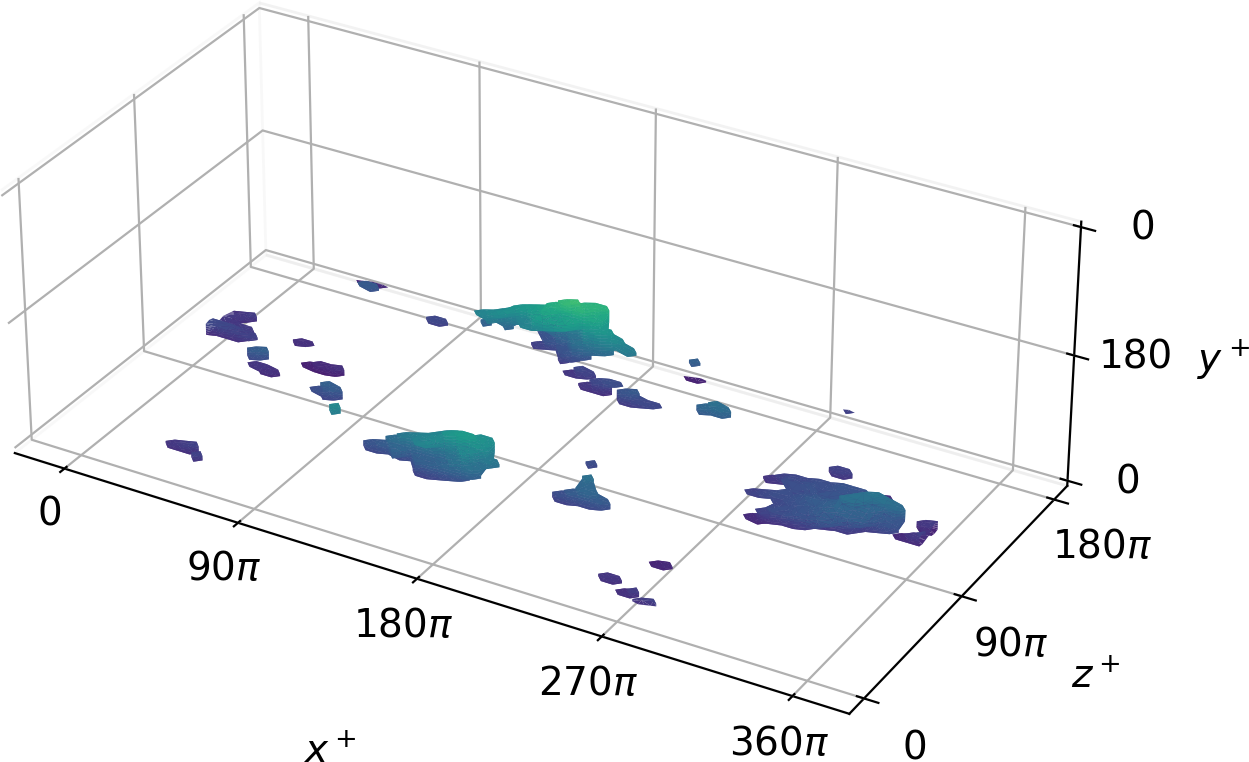}} & 
  \raisebox{-.5\height}{\includegraphics[width=0.42\linewidth]{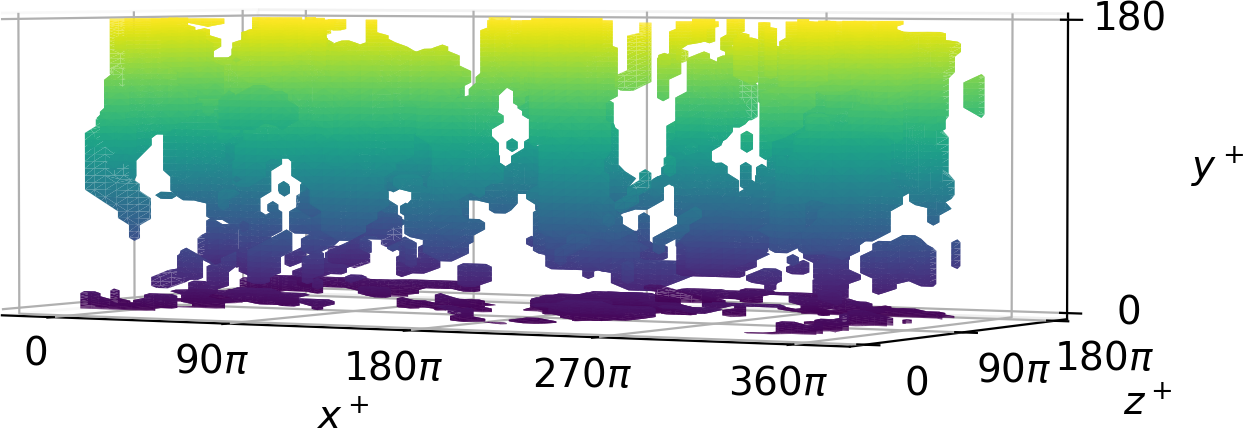}} \\

  DRL Q events &
  \raisebox{-.5\height}{\includegraphics[width=0.42\linewidth]{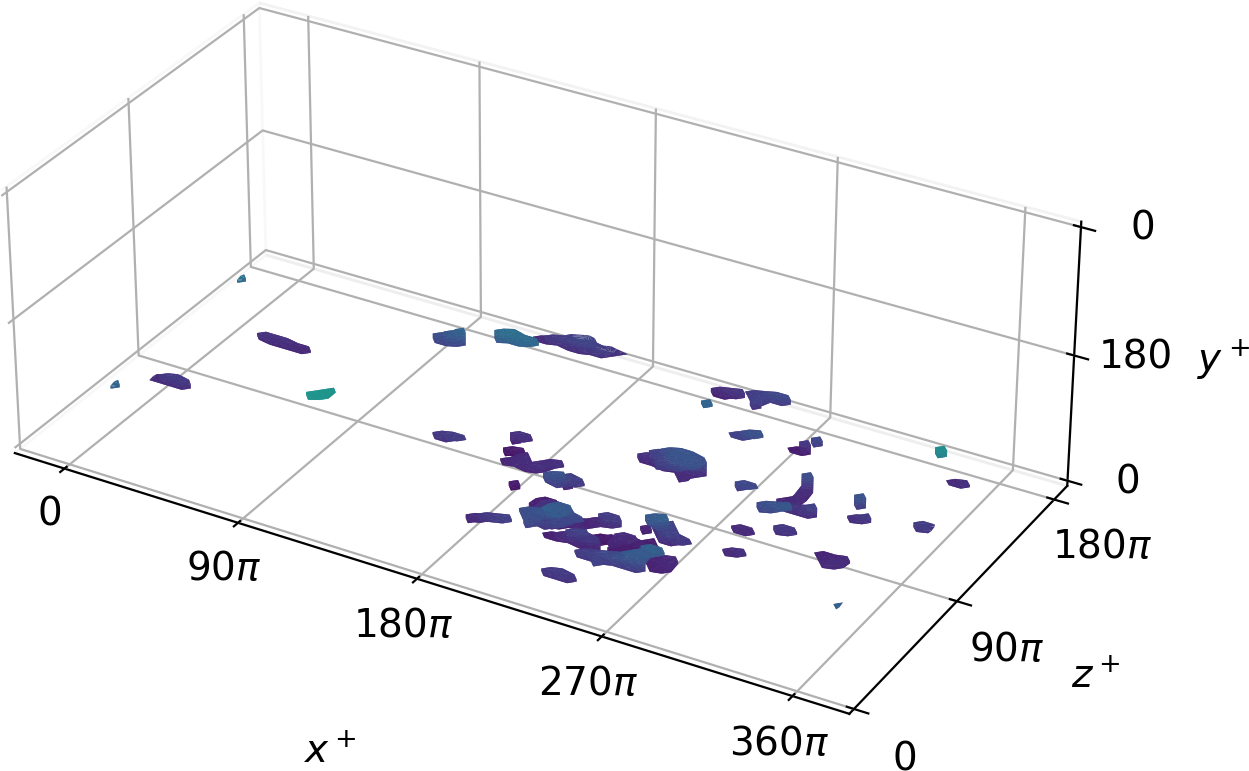}} & 
  \raisebox{-.5\height}{\includegraphics[width=0.42\linewidth]{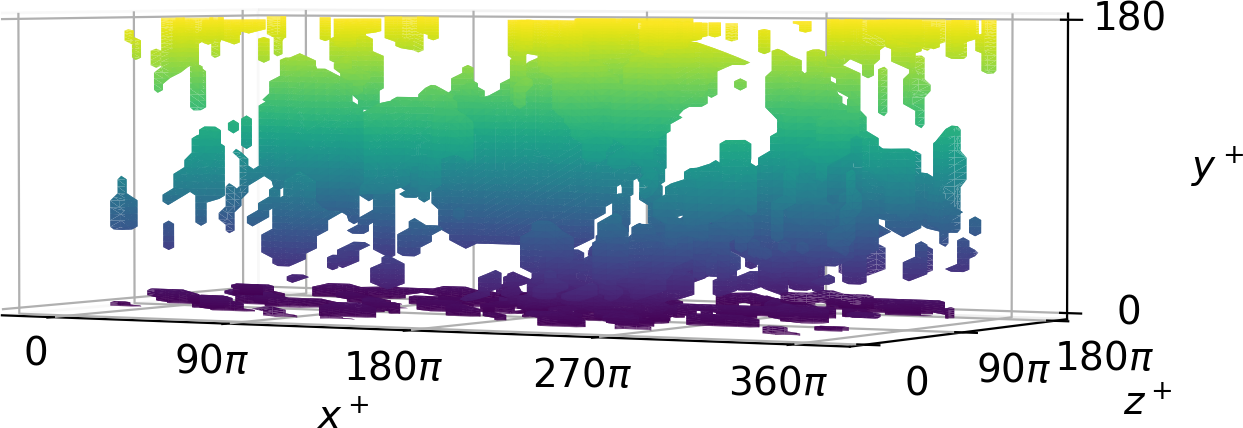}} \\

  DRL streaks &
  \raisebox{-.5\height}{\includegraphics[width=0.42\linewidth]{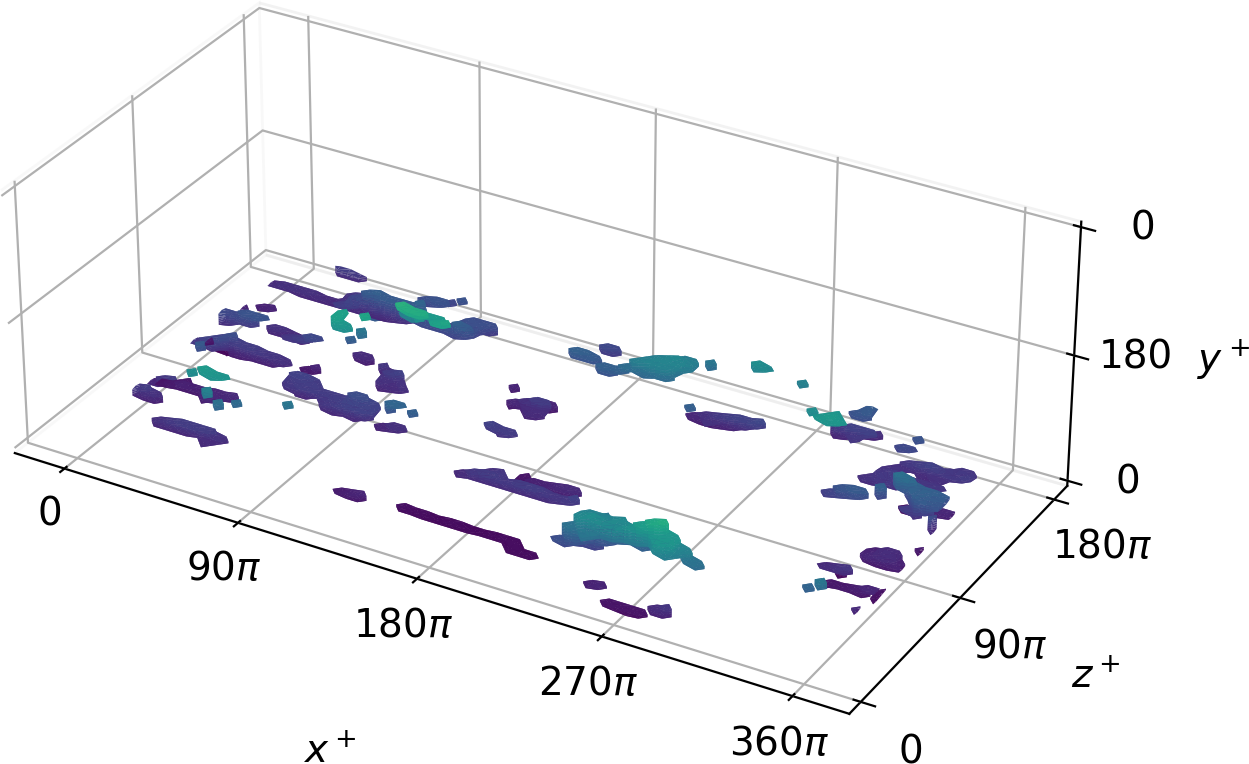}} & 
  \raisebox{-.5\height}{\includegraphics[width=0.42\linewidth]{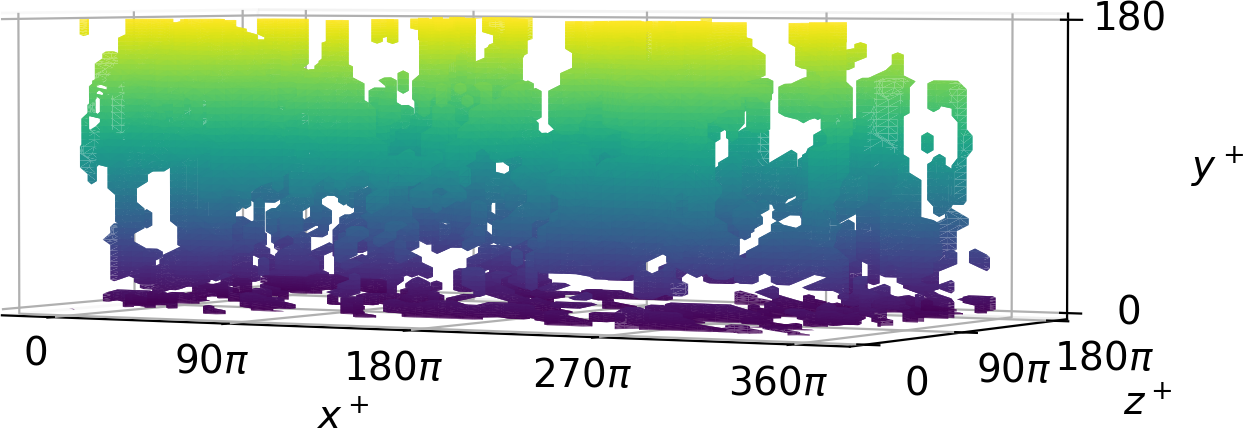}} \\

  DRL SHAP &
  \raisebox{-.5\height}{\includegraphics[width=0.42\linewidth]{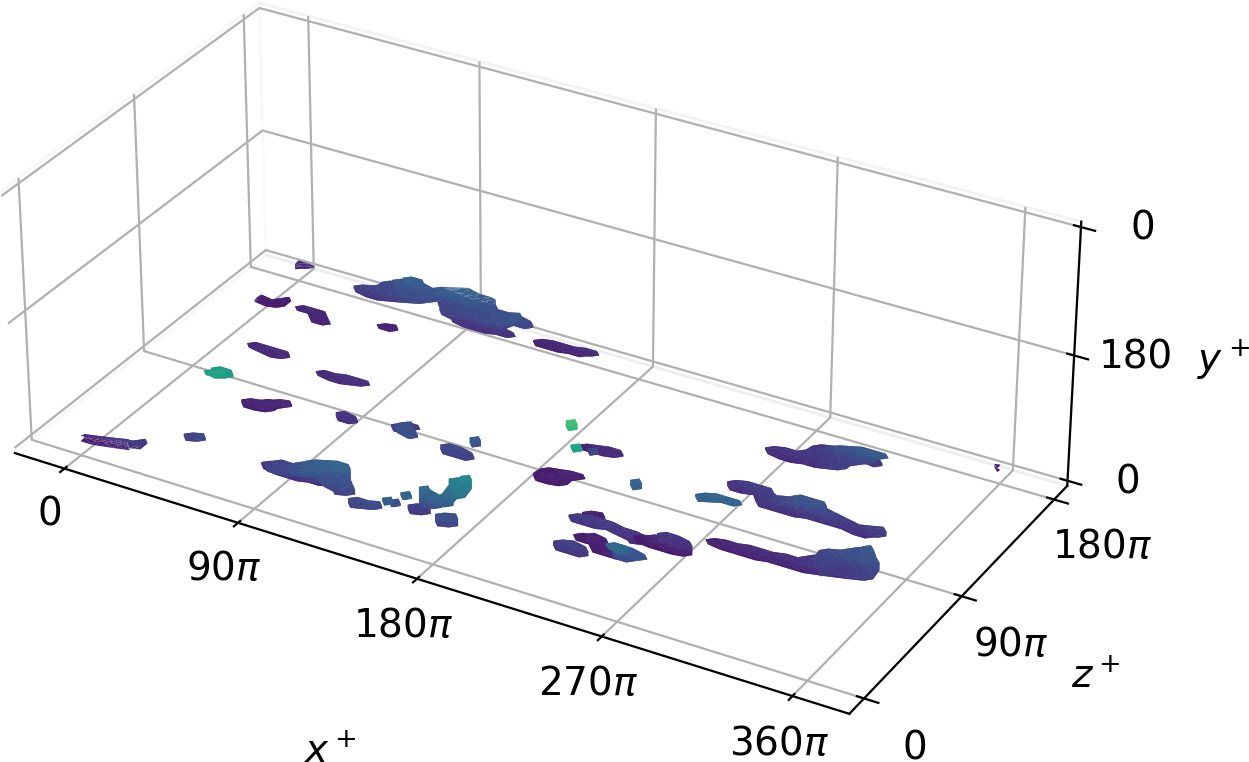}} & 
  \raisebox{-.5\height}{\includegraphics[width=0.42\linewidth]{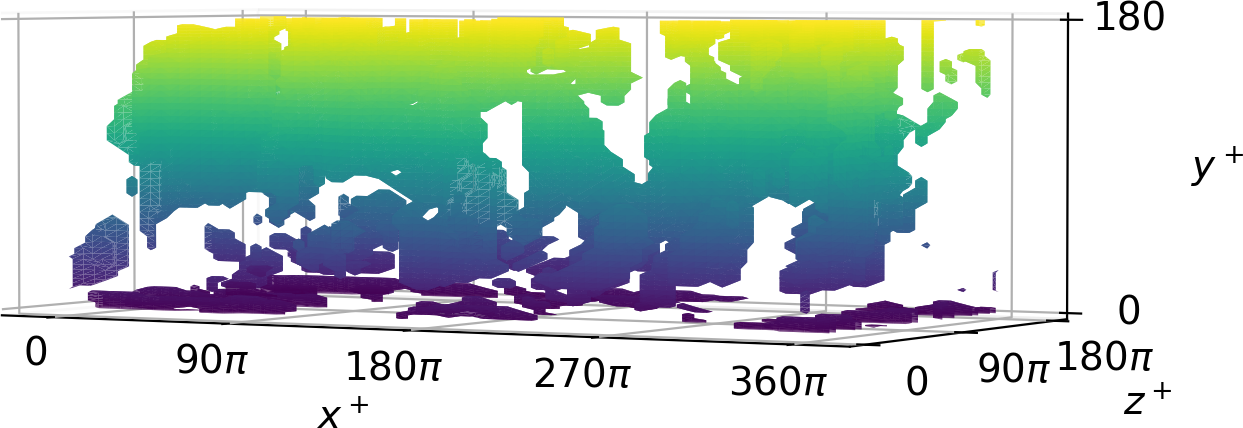}}
\end{tabular}
\caption{\textbf{Sample 3D visualizations of coherent structures in the LCC.} (Left) Streaks identified for the different control strategies once the flow has adapted to the control, i.e. $t^+>1300$. (Right) Idem for Q events. Isosurfaces are chosen following the percolation analysis in \cite{cremades2024classically} and only the outmost contour of the structures is shown. The structures are coloured based on the distance from the wall, flow from left to right.}
\label{fig:3D}
\end{figure}

\begin{figure}
    \centering
    \includegraphics[width=1.0\linewidth]{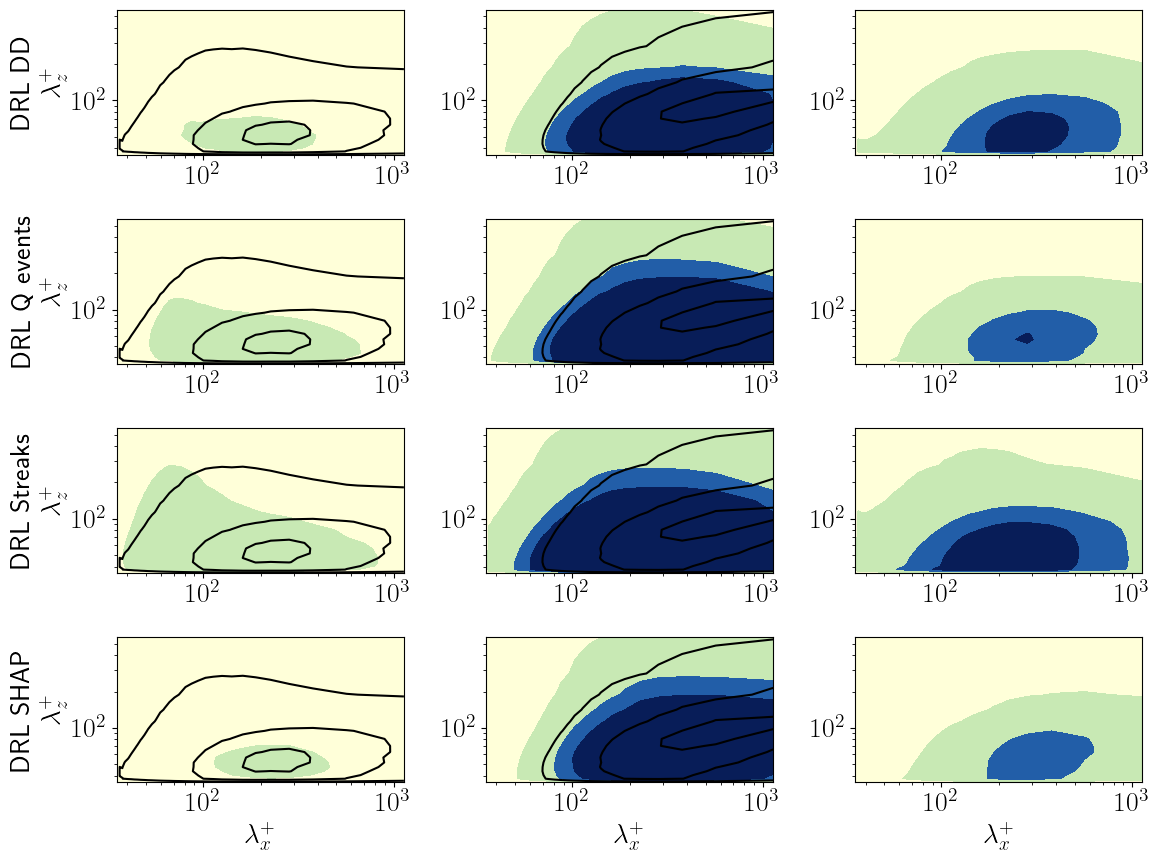}
    \caption{\textbf{Two-dimensional premultiplied spectra in the LCC}. Results are averaged over 50 initial conditions (as this can be deemed sufficent) used for policy evaluation. (Left) Two-dimensional premultiplied spectrum of the wall-normal velocity component on the sensing plane $v'(t^+=100,x,y^+=15,z)$, (middle) two-dimensional premultiplied spectrum of the streamwise velocity component on the sensing plane $u'(t^+=100,x,y^+=15,z)$, (right) two-dimensional premultiplied spectrum of the control  $v'(t^+=100,x,y^+=0,z)$. Black lines indicate the uncontrolled case and the shown levels are $[10,50,90] \%$ of the uncontrolled maximum in all cases.}
    \label{fig:spectra}
\end{figure}

\begin{figure}
    \centering
    \includegraphics[width=1.0\linewidth]{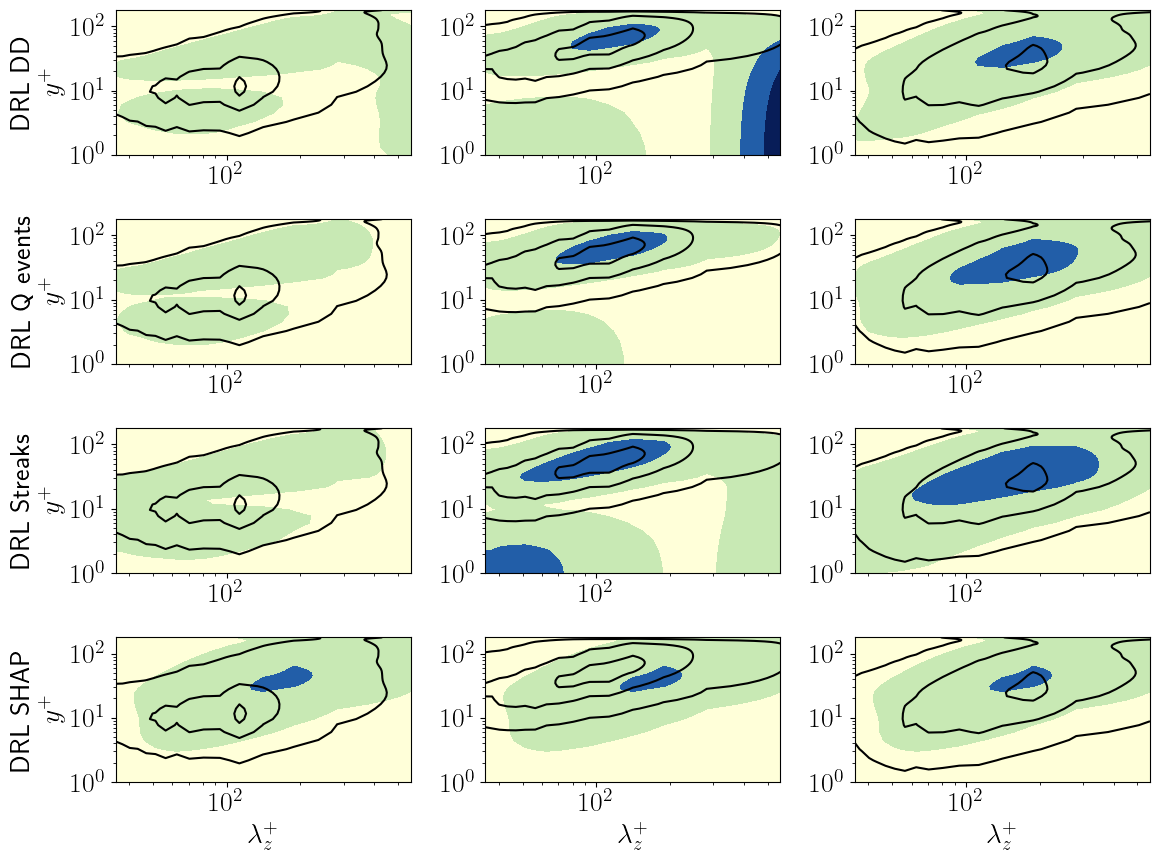}
    \caption{\textbf{One-dimensional premultiplied spanwise spectra in the LCC}. Results are averaged over 50 initial conditions (as this can be deemed sufficent) used for policy evaluation. (Left) One-dimensional spectrum of the streamwise velocity component as a function of the wall-normal direction, (middle) idem for the wall-normal velocity component, (right) idem for the spanwise velocity component. Black lines indicate the uncontrolled case and the shown levels are $[10,50,90] \%$ of the uncontrolled maximum in all cases.}
    \label{fig:spectra_yz}
\end{figure}

The statistical robustness of our computations is assessed in Fig.~\ref{fig:quant_policy}. In Fig.~\ref{fig:quant_policy} (left) we show the average value of the drag reduction for each control strategy with respect to the number of initial conditions considered. It can be observed  that beyond 75 initial conditions the resulting drag reduction is converged. In Fig.~\ref{fig:quant_policy} (right), we show a histogram of the drag reduction resulting from each individual IC. This figure clearly illustrates the differences in mean value and standard deviation for the drag reduction corresponding to each control strategy. The SHAP-based control exhibits a mean drag reduction of $33.7\%$ with a standard deviation of $\pm4.3\%$, the direct-drag reduction control shows a mean drag reduction of $31.9\%$ with a standard deviation of $\pm5.5\%$, the Q-events-based control presents a mean drag reduction of $26.9\%$ with a standard deviation of $\pm6.2\%$, the streak-based control exhibits  a mean drag reduction of $20.9\%$ with a standard deviation of $\pm7.1\%$, and the baseline opposition control presents a mean drag reduction of $22.9\%$ with a standard deviation of $\pm3.61\%$. This shows that the SHAP-based control is not only the policy that performs best in terms of drag reduction, but it is also the one that is most likely to do well, i.e. has the lowest standard deviation of all the reinforcement-learning-discovered policies.

We have also considered (i) promoting the negative Q events overall in the domain, i.e. $\gamma = \int_{V}uv$, such that $r_t=\frac{\gamma}{\gamma_{\text{uncontrolled}}}$, (ii) keeping the sign in the Q events, i.e. $\gamma = \int_{V}uv$, such that $r_t=1-\frac{\gamma}{\gamma_{\text{uncontrolled}}}$. Exploratory evaluation of these policies in a single IC in the SCC yields a drag increase of over 200$\%$ in the former case and a moderate DR of $8\%$ in the latter (much lower than any of the policies reported in the main text). We therefore consider that these choices of reward do not provide a viable alternative for the proxies defined in the text to improve turbulence control using coherent structures.

Sample visualizations the DRL-discovered control strategies at $t^+=100$ starting from the same initial condition are shown in Fig.~\ref{fig:planes_100}. The choice of showing this particular time is such that the evolution of the controlled flow is still sufficiently similar to allow for a direct comparison between the different policies. The figure shows the inputs received by the agent corresponding to the state (first two columns) as well as the selected action (last column) for each DRL-discovered control strategy. Similarly to the opposition control, the DD-based and streak-based policies tend to increase the blowing at the wall in those regions in which the wall-normal velocity is negative at the sensing plane. The Q-event-based policy introduces suction in the regions of low streamwise velocity (ejections) to avoid the lifting effect of the structures and exhibits concentrated blowing below high-velocity regions (sweeps), deflecting them from the wall. Finally, the SHAP-based policy exhibits suction in the regions in which the wall-normal velocity is positive. In addition, the causal implications of the SHAP are connected with the fact that the control blows slightly before the low-velocity region. Fig.~\ref{fig:planes_1500} shows sample visualizations at $t^+=1500$, i.e. once the flow has adapted to the control. Our results indicate that for the DD-policy and the SHAP-based policy, most of the wall-normal and streamwise perturbations are suppressed. In contrast, it can be observed that in the Q-event-based control, several small-scale perturbations remain. 

Fig.~\ref{fig:3D} shows sample three-dimensional visualizations for streaks as well as Q events for all control strategies in the LCC. The structures are identified following~\cite{cremades2024classically} and the snapshots are taken once the flow has adapted to the control, i.e. $t^+>1300$. All control strategies exhibit a significant reduction of the streaks, where the Q-events-based control and the SHAP-based control are the control strategies showing the largest streak reduction. In contrast, the DD-based control exhibits larger streaks located farther away from the wall and the streak-based control exhibits a greater number of structures. Similarly, Q events are most significantly reduced in the Q-event-based control and the SHAP-based control, noting that the region where these structures are reduced is mainly located at $y^+\sim 15$. Note that this is the region where the most intense Q events are present~\cite{cremades2024identifying}, and therefore it makes sense that this is the focus of the various control strategies. These results are in agreement with Fig.~2 in the main body of the paper, which shows various performance metrics for our control strategies.

Fig.~\ref{fig:spectra} shows the two-dimensional premultiplied spectra for the streamwise, wall-normal and spanwise fluctuations corresponding to the different control strategies on the sensing plane as well as the control plane. Darker colors indicate higher energetic content, while all spectra are normalized with respect to the uncontrolled case. We observe that the energetic content of all scales of the wall-normal velocity fluctuations is significantly reduced, being the reduction most notable for the DD-based control and the SHAP-based control. In all cases we can observe that the energy shifts toward smaller scales in $x$ ($\lambda_x^+\approx 100$) when compared to the uncontrolled case, and that at the same time, the energy is distributed over a wider range of scales in $z$. The spectrum of the streamwise fluctuations exhibits a good agreement with the peak present in the uncontrolled case. We observe a greater energetic content at the larger scales in $x$ in the spectra of all control strategies, while there is little change in scales in the $z$ direction. The streak-based control stands out here since it exhibits the largest range of scales in $x$ of all control strategies. Analyzing the control plane, we observe that there is an energy peak that correlates well with the uncontrolled spectrum of the wall-normal fluctuations ($\lambda_x^+\approx 300$ and $\lambda_z^+\approx 50$). This peak is present in all control strategies, although the energy content and extension of the scales is quite different. The SHAP-based control exhibits the least energetic peak in the spectrum, followed by the Q-event-based control. At the other end we find the streak-based control, whose energy is distributed over a wide range of scales. These results indicate that the SHAP-based control targets structures in a more specific way than any of the other methods.

Fig.~\ref{fig:spectra_yz} shows the one-dimensional premultiplied spanwise spectra as a function of the wall-normal coordinate corresponding to the various control strategies and darker colors indicate higher energetic content, while spectra are normalized with respect to the uncontrolled case. The one-dimensional spectra in the streamwise direction is significantly affected by the various control strategies: the DD-based control, Q-event-based control, and the streak-based control exhibit two energy peaks: one located at $y^+\approx 6$ and another one located at $y^+\approx 20$ extending into $y^+\approx 100$, both peaks spanning a similar range of spanwise scales between $\lambda_z^+\approx 40$  and $\lambda_z^+\approx 100$. In contrast, the SHAP-based control has only one energy peak, which is located farther away from the wall than the uncontrolled case. Moreover, the maximum energy content in the spanwise scales is concentrated within wavelengths about twice as large as those in all other control strategies ($\lambda_z^+\approx 200$ for the SHAP-based control vs. $\lambda_z^+\approx 100$ all other control strategies). In the wall-normal direction, we can again observe that SHAP-based control modifies the uncontrolled spectrum the least, mainly pushing the energy content towards larger lengthscales (note the relative correspondence of the energy peaks in the streamwise and wall-normal velocity components). All other control strategies exhibit a new energy peak at the small scales $\lambda_z^+\approx 30$ located very close to the wall. Furthermore, the DD-based control and the streak-based control exhibit one additional peak for large spanwise wavelengths $\lambda_z^+\approx 500$ which extends throughout the channel. Note that the spectrum of the spanwise velocity fluctuations is minimally affected by the control. Interestingly, our results suggest that the SHAP-based policy is capable of controlling turbulence without greatly affecting the scales present in the flow, while the control strategies based con classically studied coherent structures give rise to a wide range of spatial scales in the flow.

\section*{A Navigation Problem}

\begin{figure}
    \centering
    \includegraphics[width=0.39\linewidth]{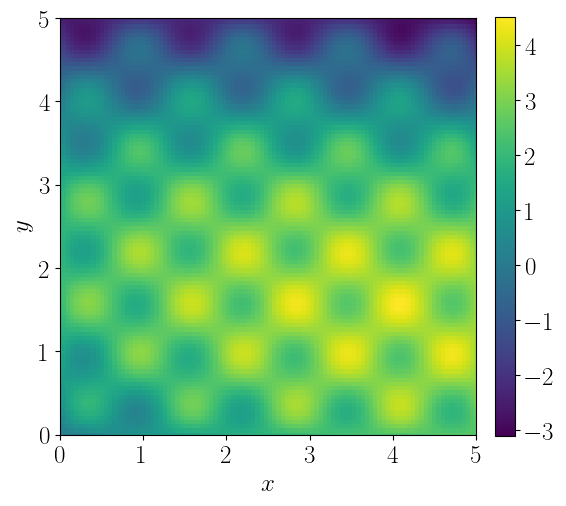}
    \includegraphics[width=0.39\linewidth]{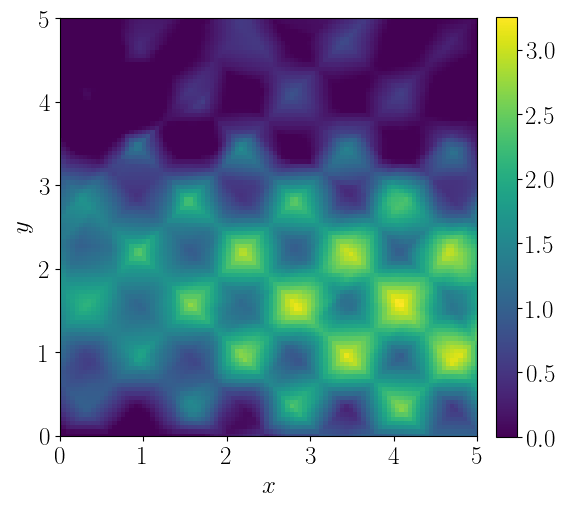}
    \includegraphics[width=0.39\linewidth]{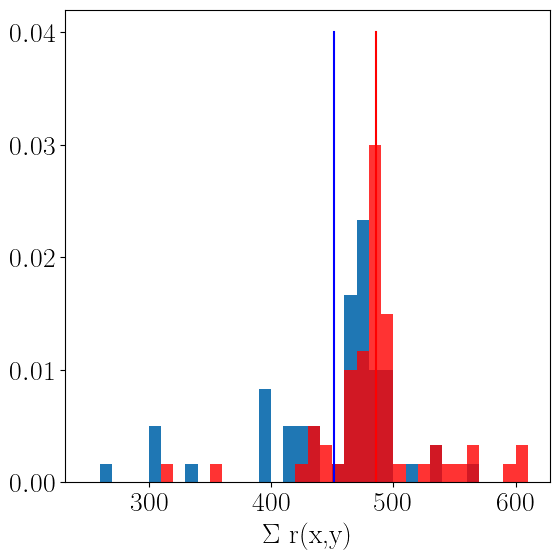}
    \includegraphics[width=0.39\linewidth]{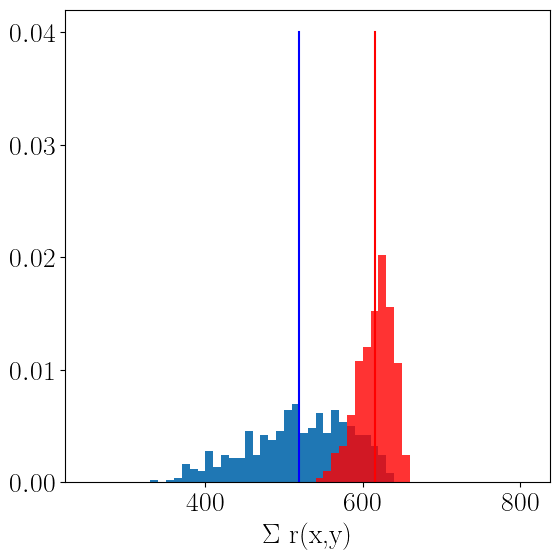}
    \caption{\textbf{Navigation problem}. (Top left) Contours of the reward given by $r(x,y)$ for the grid problem. (Top right) Contours of the reward given by $r_{\text{SHAP}}$, the regions that`positively contribute to $r(x,y)$ are highlighted. (Bottom left) Histogram for the evaluation on one initial condition for 60 different trained models. This corresponds to the numerical test (i) described in the text. (Bottom right) Histogram for the evaluation of the best model trained on each reward for 600 different initial conditions. Red denotes the model trained on a SHAP-based reward and blue the model trained on a direct reward. Evaluation is always done on the direct reward. Vertical lines show mean values across models/runs.}
    \label{fig:navigation_model} 
\end{figure}

The high dimensionality of the turbulent flow considered here and its high spatio-temporal complexity make it difficult to provide a simple explanation for the better performance of the SHAP-based policy with respect to its counterparts.  This consideration motivates the adoption of a two-dimensional navigation problem on which our concepts can be illustrated. We choose a two-dimensional problem due to its simplicity and low computational cost, since numerical simulations even in the SCC imply a significant computational load. We  can then draw qualitative analogies to the turbulent channel flow directly from our navigation problem. This two-dimensional benchmark presents the ideal conceptual testbed since it allows for a complete visualization of the full state space, a quick training of reinforcement learning models, and many-query testing. The landscape of our two-dimensional navigation problem reads
\begin{equation}
    r(x,y):= x+2y-\frac{xy}{8}-\frac{x^2}{10}-\frac{y^2}{2}+\sin(5x)\sin(5y),
\end{equation}
where $(x,y)$ denote the two features of a full state. We can formulate an analogous optimization problem to that of the turbulent channel: Consider an agent that tries to maximize $r(x,y)$ for an episode of $N=150$ steps, taking steps of size $\Delta=0.2$ on a grid of size $[0,5]\times [0,5]$, and has 5 possible actions, move up, move down, move left, move right, or stay. The analogy with the turbulence control problem is clear: $x,y$ represent the observations, $r$ is analogous to the drag reduction, and the actions represent the possible blowing and suction. The number of steps allows it to traverse the grid several times, but is finite, akin to the limited number of training episodes available in the turbulent channel. We note that a key difference with the turbulent channel is that no evolution laws are considered here, such that the $r(x,y)$ does not change as an agent takes steps.

We pose two reinforcement learning problems in this set-up: given the current state $(x,y)$ find the optimal policy to maximize (i) $r(x,y)$ and (ii) a reward $r_{\text{SHAP}}(x,y):=\max{(0,\text{SHAP}_x)}+\max{(0,\text{SHAP}_y)}$, i.e. the SHAP values that contribute positively to $r(x,y)$. The resulting policies will be evaluated on the original $r(x,y)$. This is analogous to our turbulent problem, since we set-up a DRL framework to optimize, $r(x,y)$ (the drag reduction) and another reward based on the SHAP values, but the performance on inference is evaluated on $r(x,y)$. To emphasize the analogy with the turbulent flow problem, we apply the same framework to calculate the SHAP values: we train an MLP to predict $r(x,y)$ from an $(x,y)$ pair, and then we calculate the SHAP values for this MLP. The results are shown in Fig.~\ref{fig:navigation_model}. We show the values of the function $r(x,y)$ in Fig.~\ref{fig:navigation_model} (top left), and the SHAP values obtained through our MLP in Fig.~\ref{fig:navigation_model} (top right). This figure shows that $r_{\text{SHAP}}$ only highlights a particular region of the $(x,y)$ space, which effectively can give the DRL training.

We then perform two numerical experiments: (i) We train 60 different models for 3500 episodes and evaluate on a single initial condition. (ii) We consider the best performing model out of test (i) and evaluate it on 600 arbitrary initial conditions. The results for test (i)  are shown in Fig.~\ref{fig:navigation_model} (bottom left) where we observe that models trained on a SHAP-based reward outperform on average models trained directly on the reward. Moreover, training on a SHAP-based reward is shown to result in more models with significantly  above average performance. The results for test (ii) are shown in Fig.~\ref{fig:navigation_model} (bottom right), where we observe that the best model trained on $r_{\text{SHAP}}(x,y) $ significantly outperforms the model trained directly on $r(x,y)$, even when evaluated on $r(x,y)$. This is observed both in terms of the mean values, but also in terms of standard deviation, where the best model trained on $r_{\text{SHAP}}(x,y) $ has a lower standard deviation. The observations for the navigation problem presented here are in qualitative agreement with those observations for the turbulent channel flow. These results illustrate how in a complex optimisation landscape with many local optima, directly reducing $r(x,y)$ can easily result in policies stuck in one of these optima, while appropriately shaping the reward with the SHAP values can help to guide the training by modifying the space over which the DRL is optimizing
.
\section*{\MB{Sensitivity analysis to sensing plane and time interval for SHAP}}
\MB{
In order to ensure the robustness of the results for the SHAP-based control, we perform a sensitivity study on the sensing plane $y^+$ and the time horizon for the SHAP calculation $\Delta t^+$. Besides the original sensing plane at  $y^+=15$, we also study planes located at  $y^+=30$ and $y^+=50$. We train a DRL model with observations at each $y^+$ for SHAP values and direct drag reduction (the two most promising control strategies), using $\Delta t^+$ for the computation of the SHAP values. We evaluate opposition control at these two $y^+$ locations as shown in figure \ref{fig:shap_comp}.

\begin{figure}
    \centering
    \includegraphics[width=0.49\linewidth]{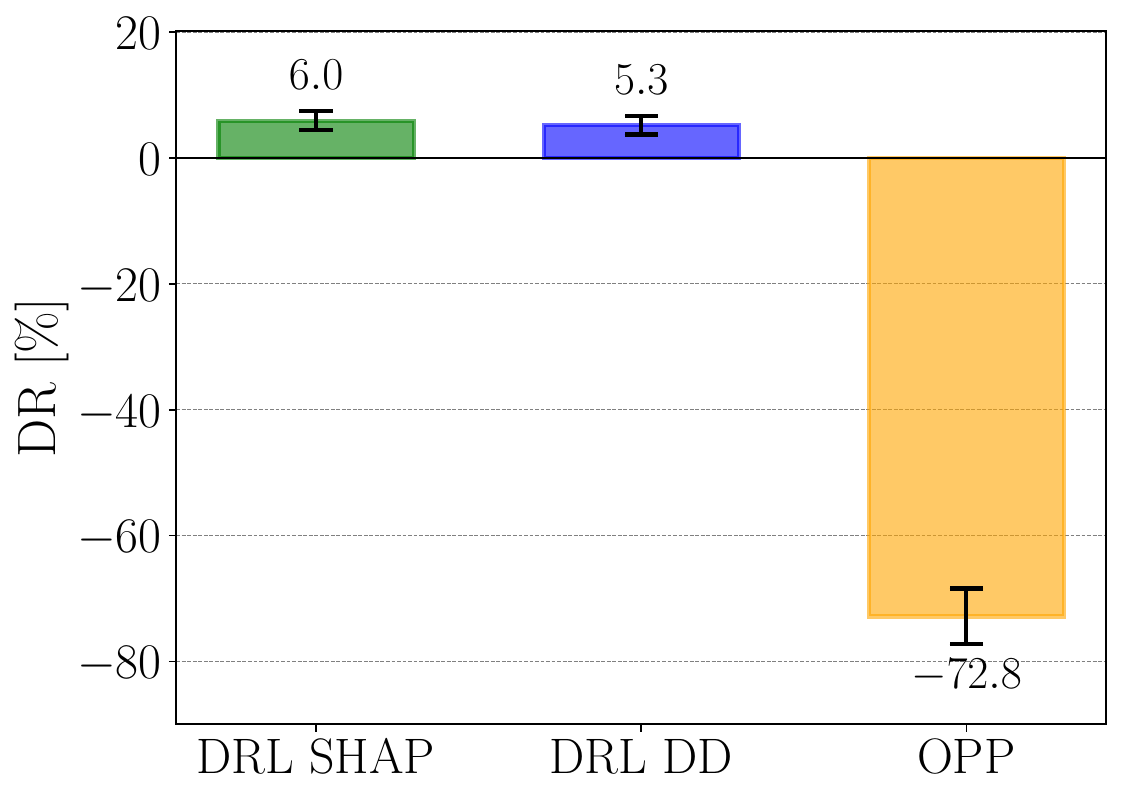}
    \includegraphics[width=0.49\linewidth]{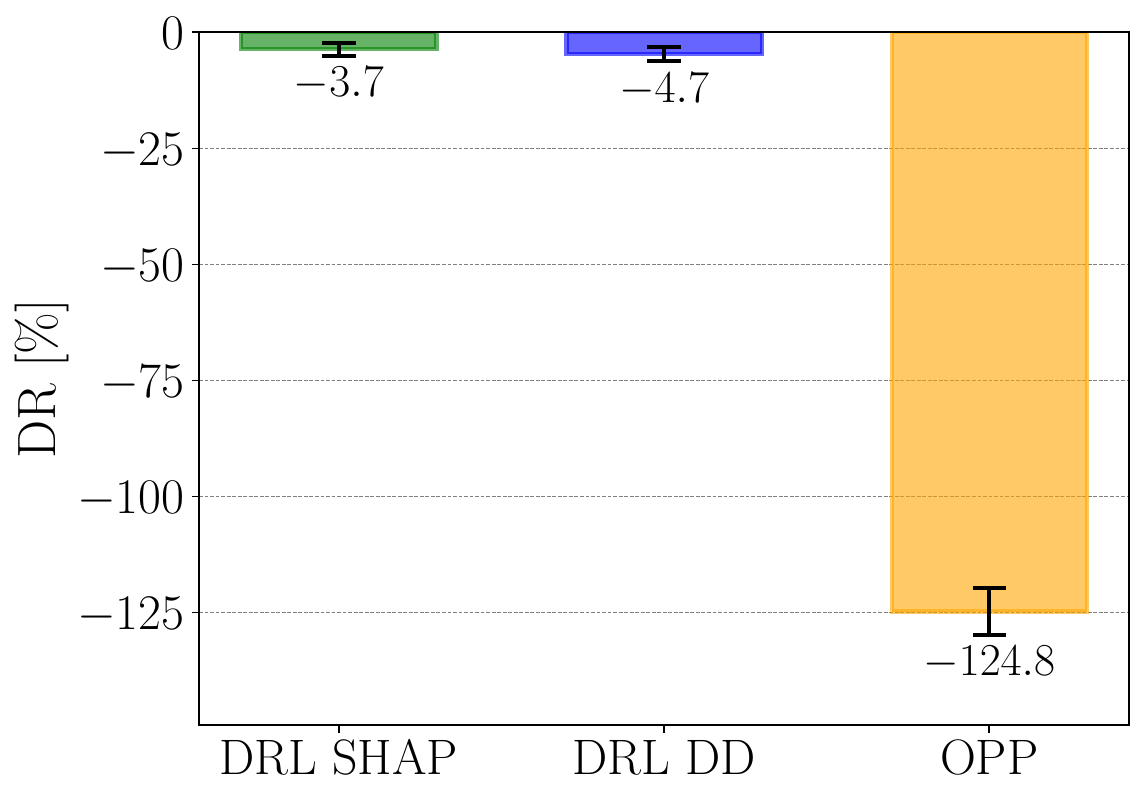}
    \caption{\MB{\textbf{ Drag reduction for varying sensing plane.}  (Left) Drag reduction for the SHAP-based policy and direct-drag-reduction policy for the best model trained with observations at $y^+=30$, opposition control is evaluated as a comparison. (Right) Same quantities for observations at  $y^+=50$.}}
    \label{fig:var_tplus} 
\end{figure}

All the models experience an expected degradation of the drag reduction at $y^+=30$ and $y^+=50$ compared to $y^+=15$, as previously documented in the literature \cite{vinuesainfluence}. Note however, that in all cases, the SHAP-based control degrades more gracefully, outperforming direct drag reduction. This study provides evidence that the SHAP-based policy is the best performing control strategy, regardless of the chosen $y^+$.

When it comes to the time horizon used to calculate the  SHAP values, we will consider three different $\Delta t^+=\{ 5,7,10\}$ for the SHAP calculation. Let us first compare the SHAP fields at these objective times in the small and large channels as shown in figure \ref{fig:shap_comp}. 

\begin{figure}
    \centering
    \includegraphics[width=0.32\linewidth]{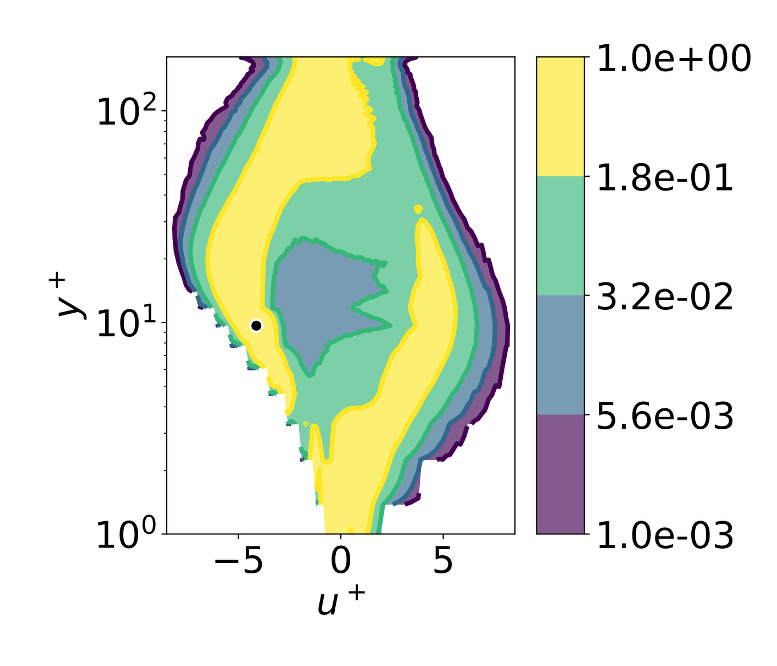}
    \includegraphics[width=0.32\linewidth]{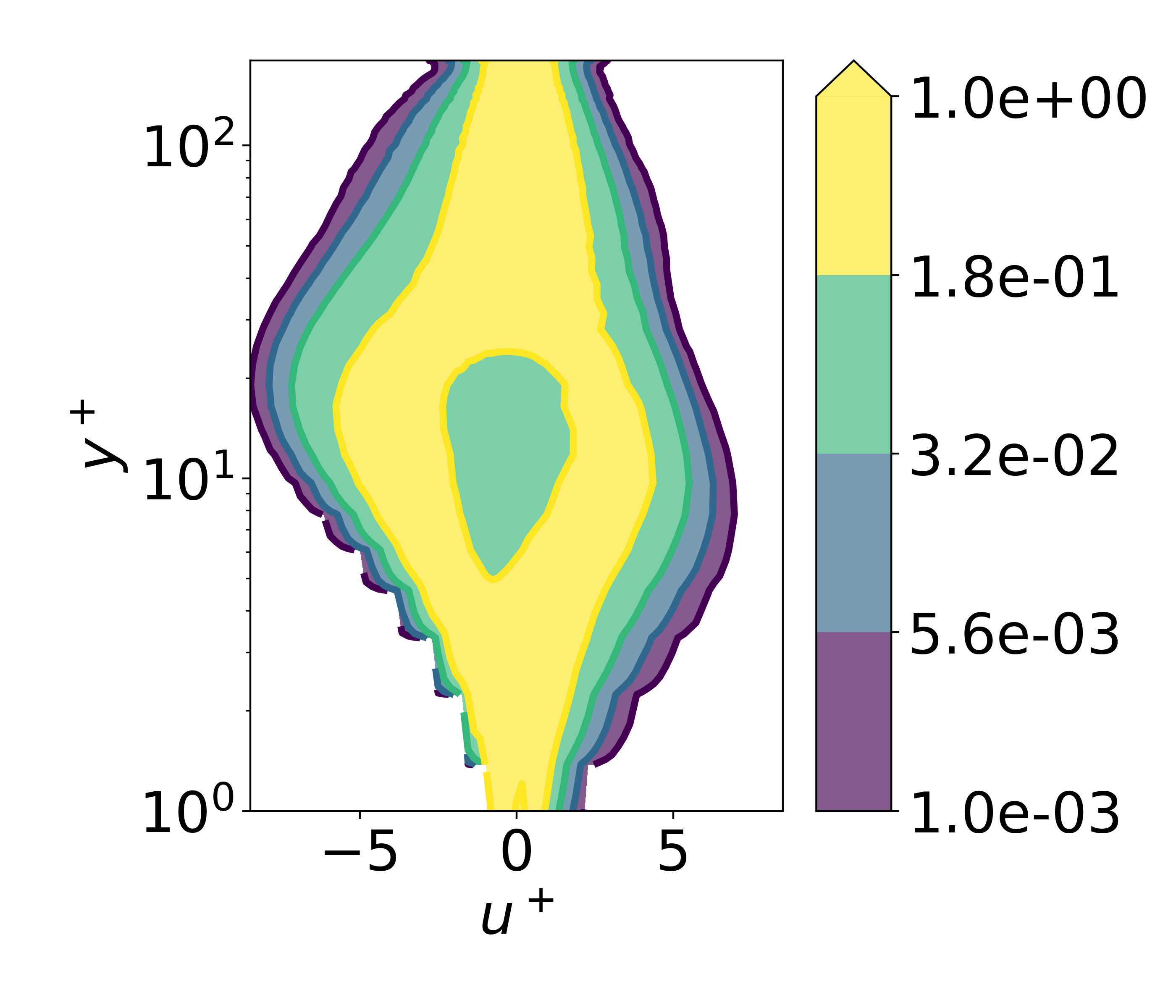}    \includegraphics[width=0.32\linewidth]{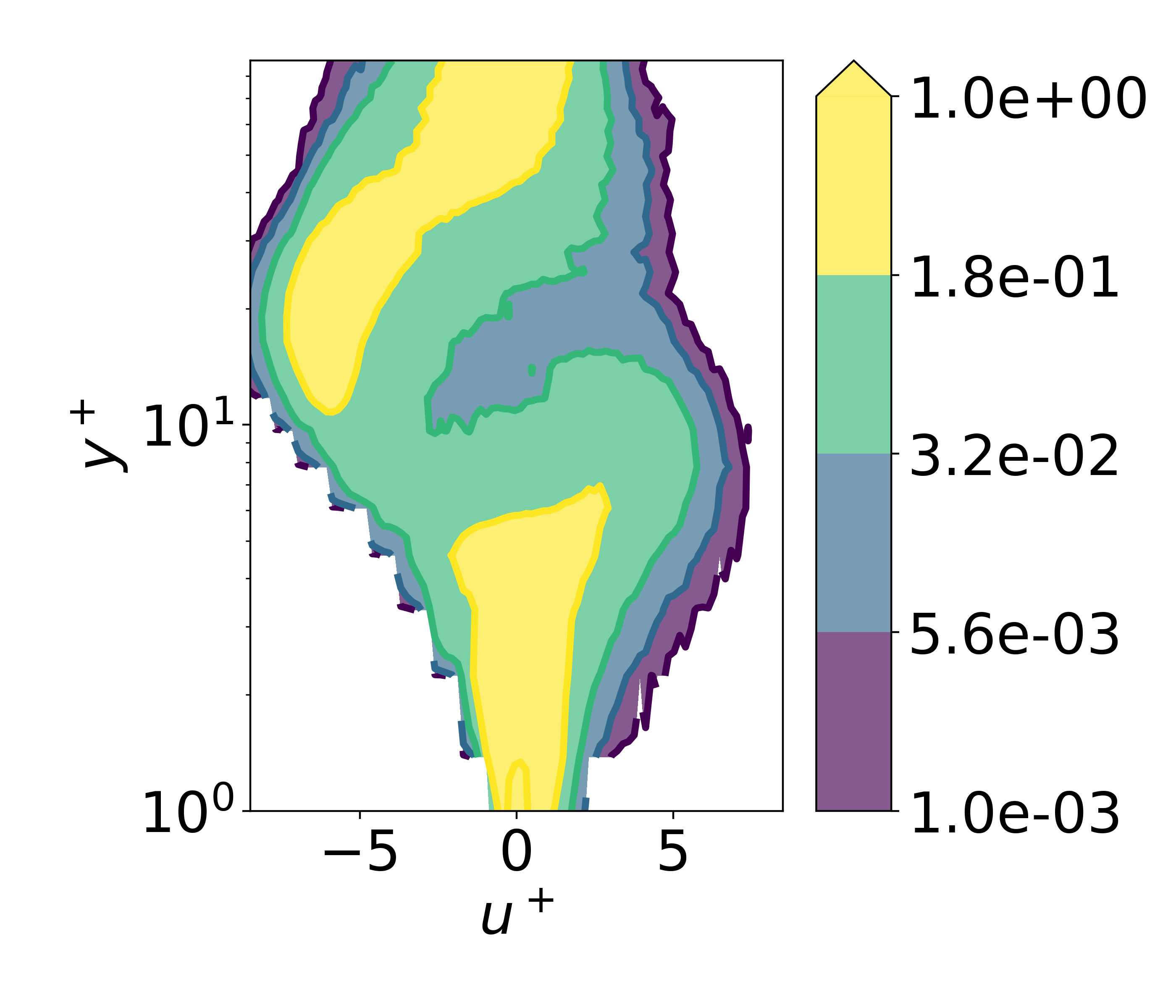}
    \includegraphics[width=0.32\linewidth]{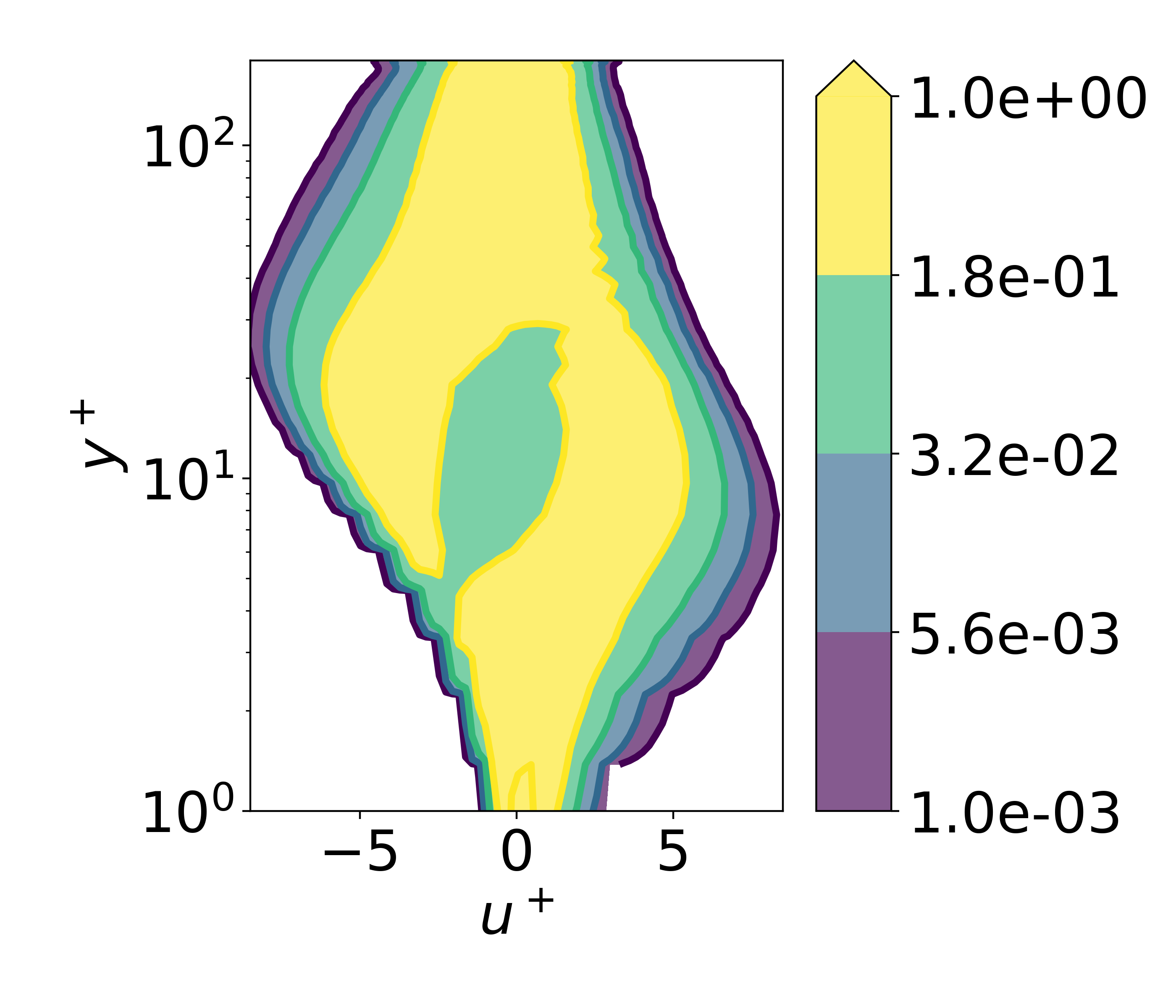}
    \includegraphics[width=0.32\linewidth]{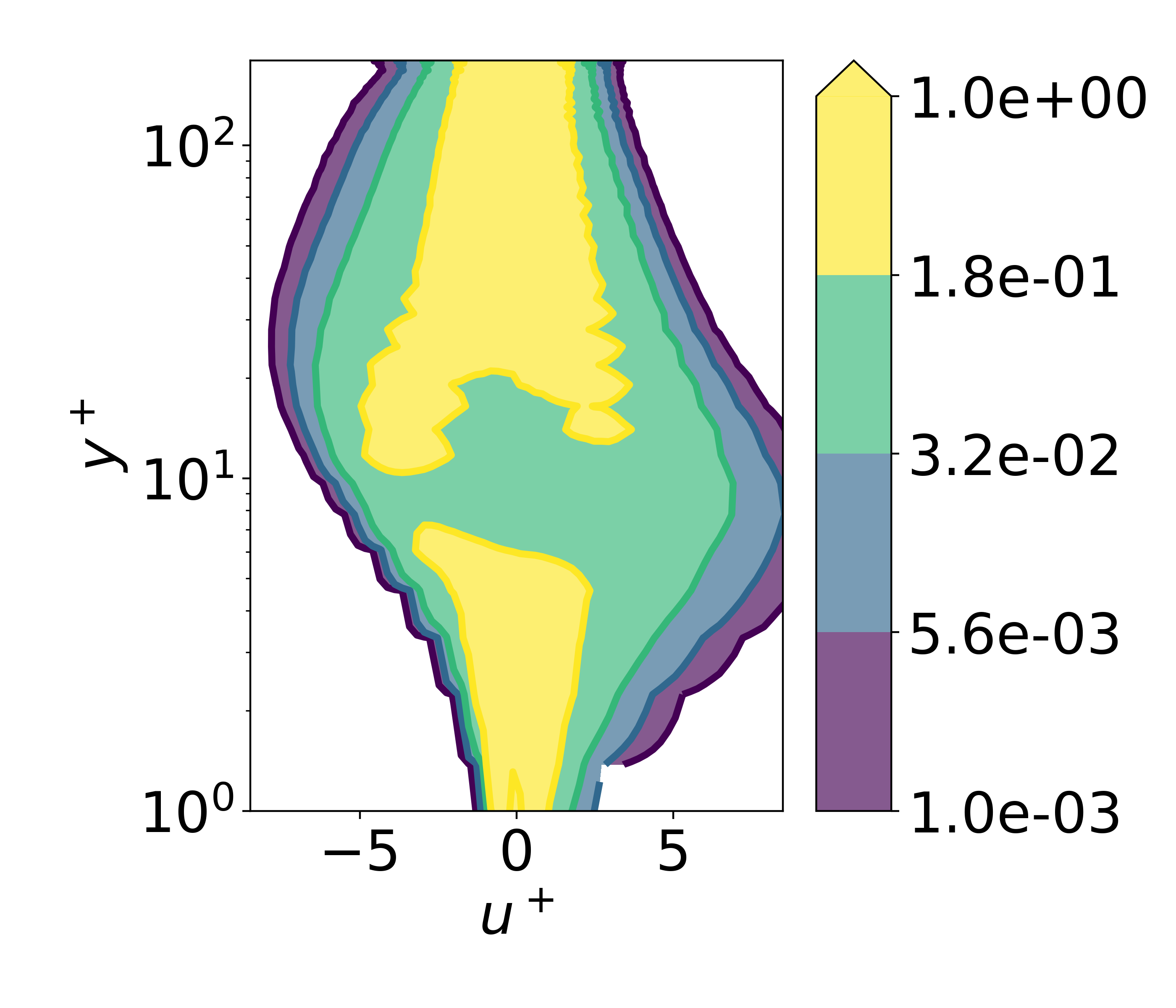}
    \includegraphics[width=0.32\linewidth]{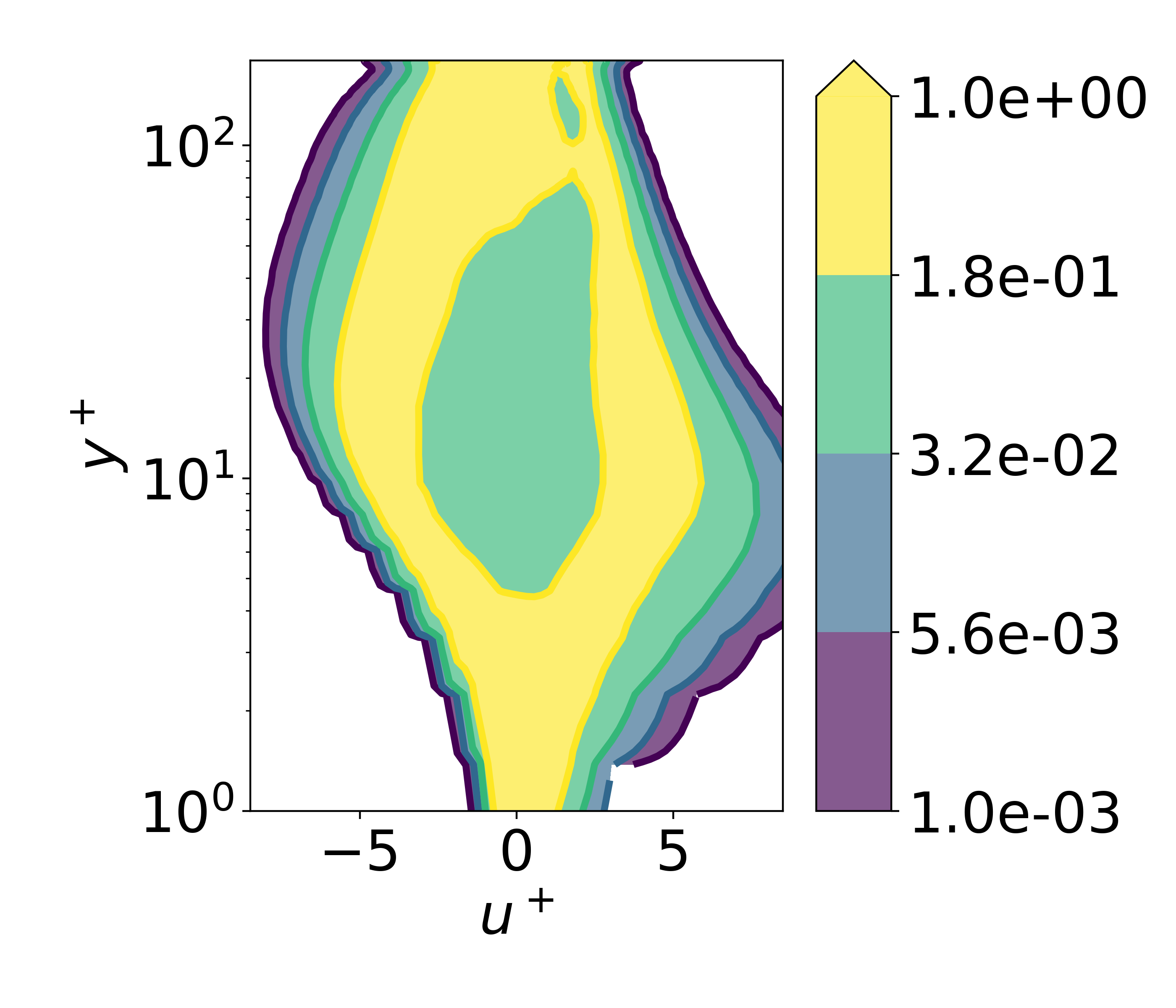}
    \caption{\MB{\textbf{SHAP values in the streamwise direction comparison.} (Top) small channel: the SHAP-values in the streamwise direction as a function of the wall distance for $\Delta t^+=5$ (left), $\Delta t^+=7$ (middle), $\Delta t^+=10$ (right). (Bottom) Idem for the large-channel quantities averaged in $x$ and $z$.}}
    \label{fig:shap_comp} 
\end{figure}

We show in figure \ref{fig:shap_comp} that that the SHAP values for the small channel at $\Delta t^+=5$ agree well with the SHAP values for the large channel at the same $\Delta t^+=5$. We then see how this corresponding agreement degrades as $\Delta t^+$ is increased, where at $\Delta t^+=10$ the agreement between the SHAP values of the small and large channels is relatively poor. In the case where the SHAP values in the small and large channels do not agree well ($\Delta t^+=\{7,10 \}$), we perform a short transfer-learning process  from the small channel to the large channel, so that the policy trained to reduce the SHAP values in the small channel can adapt to the SHAP values in the large one. The short transfer learning is carried out by taking the satisfactorily trained agent at $\Delta t^+=5$ in the small channel and then fine tuning it in the large channel as follows: we reset only the critic, the actor is warmed-up for 16 episodes in the full channel before fine-tuning for at most 15 episodes after unfreezing. We keep the best performing model of the fine-tuning.  We do not vary any other parameter. This is done following two alternative paths: (i) Starting from the reported policy in the small channel at $\Delta t^+=5$, fine tune in the large channel for $\Delta t^+=5$. Afterwards, we do an additional fine tuning from the policy in the large channel at $\Delta t^+=5$ to the large channel at $\Delta t^+=\{7,10\}$ (ii) We perform the fine tuning from the policy corresponding to the small channel at $\Delta t^+=5$ to the large channel with a $\Delta t^+=\{5,7,10\}$, i.e. directly small to large without using the large at $\Delta t^+=5$ as an intermediate step in the fine tuning. This allows for two different routes to access the SHAP values in the large channel at the different $\Delta t^+=10$.
 
\begin{figure}
    \centering
    \includegraphics[width=1.0\linewidth]{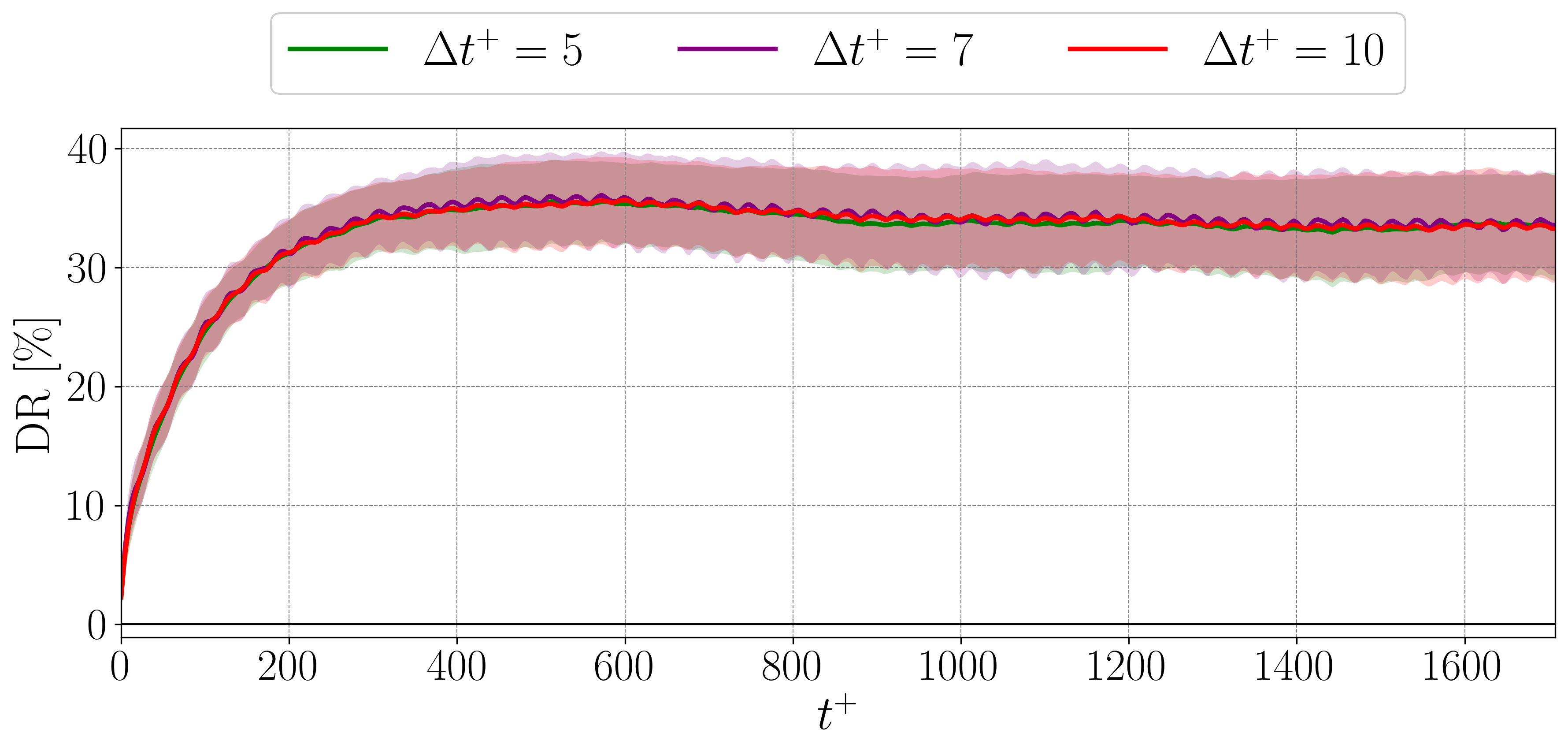}
    \includegraphics[width=1.0\linewidth]{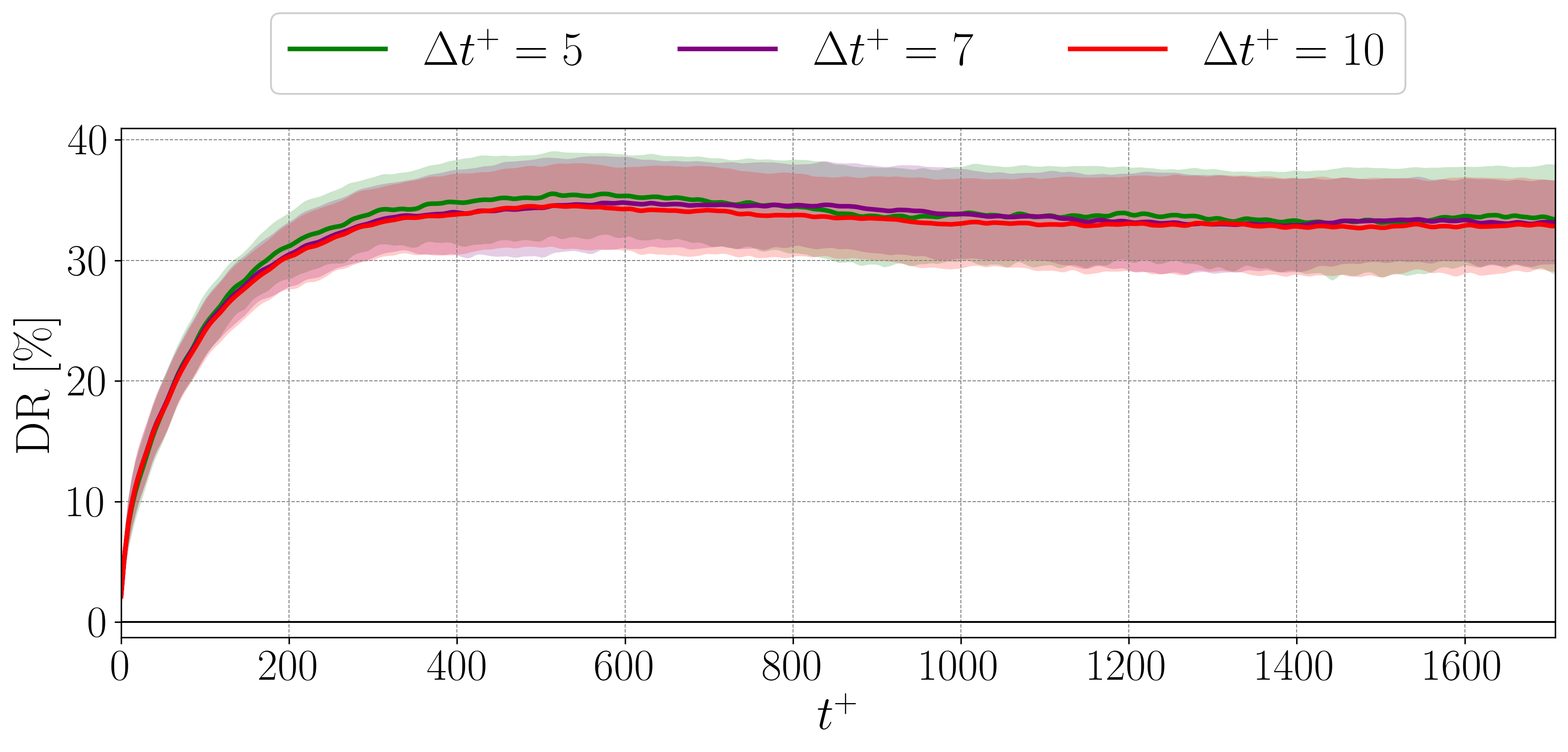}
    \caption{\MB{\textbf{Drag reduction for varying $\Delta t^+$.} (Top) Drag reduction for the SHAP-based policy fine tuned from the small channel at $\Delta t^+=5$ to the large channel at $\Delta t^+=5$ and then fine-tuned at another $\Delta t^+$. (Bottom) Drag reduction for the SHAP-based policy fine tuned from the small channel at $\Delta t^+=5$ directly to other values of $\Delta t^+$ in the large channel. The results are evaluated on 150 initial conditions in the large channel unseen during training.}}
    \label{fig:DR_dt} 
\end{figure}

Figure \ref{fig:DR_dt} shows that both transfer-learning approaches yield identical results, and also that the drag-reduction rate achieved by the DRL policy targeting SHAP values does not change with t+. In particular,  at $\Delta t^+=5$ the large-channel transfer results in $33.99\%$ DR and the small-large transfer learning results in $33.76\%$ DR; at $\Delta t^+=7$ the large-channel transfer results in $34.34\%$ DR and the small-large transfer learning results in $33.76\%$ DR; and at $\Delta t^+=10$ the large-channel training results in $34.17\%$ DR and the small-large transfer learning results in $33.33\%$ DR. The direct drag reduction policy is still worse than all the SHAP-based policies ($31.9\%$ drag reduction), a fact that showcases the robustness of our results to $\Delta t^+$ in the SHAP computation.
}

\newpage
\bibliography{sn-bibliography}% common bib file
%% if required, the content of .bbl file can be included here once bbl is generated
%%\input sn-article.bbl

\end{document}